\documentclass[preprint2]{aa}
\usepackage{txfonts}
\usepackage{graphicx,subfigure}
\usepackage{epsfig}
\usepackage{epsf}
\usepackage{lscape}
\usepackage{natbib}
\usepackage{rotating}
\usepackage{tabularx}
\usepackage{supertabular}

\newcommand{\mic}{\,{\rm \mu m} }

\newcommand{\sech}{\textrm{sech}} 

\newcommand{\Tdust}{T_{d}}           
\newcommand{\dustem}{\epsilon_{e}}   
\newcommand{\vtrans}{v_{t}}          

\newcommand{\TLS}{TLS\,}           

\begin{document}
\title{Far-infrared to millimeter astrophysical dust emission}
\subtitle{II: Comparison of the two-level systems (TLS) model with astronomical data}
\author{ D. Paradis \inst{1,2} \and J.-Ph. Bernard \inst{1,2} \and
  C. M\'eny \inst{1,2} \and V. Gromov \inst{1,2,3} }
\offprints{paradis@cesr.fr}
\institute{Universit\'e de Toulouse; UPS-OMP; IRAP; Toulouse, France
\and 
CNRS; IRAP; 9 Av. du Colonel Roche, BP 44346, F-31028 Toulouse cedex
4, France
\and
Space Research Institute, RAS, 84/32 Profsoyuznaya, 117810 Moscow,
Russia }
\date{}
  \abstract
{}
{In a previous paper we proposed a new model for the emission by
amorphous astronomical dust grains, based on solid-state physics. The
model uses a description of the disordered charge distribution (DCD)
combined with the presence of two-level systems (TLS) defects in the
amorphous solid composing the grains. The goal of this paper is to
compare this new model to astronomical observations of different
Galactic environments in the far-infrared/submillimeter, in order to derive a set of canonical model parameters to
be used as a Galactic reference to be compared to in future
Galactic and extragalactic studies.}  
{We compare the \TLS model with existing astronomical data. We consider the average emission
spectrum at high latitudes in our Galaxy as measured with FIRAS and
WMAP, as well as the emission from Galactic compact sources observed
with the Archeops balloon experiment, for which an inverse
relationship between the dust temperature and the emissivity spectral
index ($\beta$) has been proven.}  
{We show that, unlike models previously proposed that often invoke two dust components at
different temperatures, the \TLS model successfully reproduces
both the shape of the Galactic spectral energy distribution and
its evolution with temperature as observed in the Archeops data.
The best \TLS model parameters indicate a charge coherence length of
$\simeq$ 13 nm and other model parameters in  broad agreement
with expectations from laboratory studies of dust analogs.  We
conclude that the millimeter excess emission, which is often
attributed to the presence of very cold dust in the diffuse ISM, is
very likely caused solely by TLS emission in disordered amorphous
dust grains. We discuss the implications of the new model, in terms of
mass determinations from millimeter continuum observations and the
expected variations in the emissivity spectral index with
wavelength and dust temperature. The implications for analyzing of the Herschel and Planck satellite data are discussed.}
 {}


\maketitle
\section{Introduction}

Early astronomical observations in the far-infrared (FIR) and
submillimeter (submm) range have revealed the spectral energy
distribution (SED) of dust
emission towards the diffuse interstellar medium (ISM) to be consistent
with a modified black-body function of the form
\begin{equation} \label{eq:I}
I_{\nu}(\lambda)=\dustem\,B_{\nu}(\lambda,T_d) \times N_H
\end{equation}
where $\rm I_{\nu}$ is the specific intensity or sky brightness
(energy flux per unit wavelength and solid angle), $\rm B_{\nu}$ the Planck function, $\lambda$ the wavelength, $\rm T_d$ the dust
equilibrium temperature, and $\rm N_H$ the hydrogen column density.
Those measurements were found to be consistent
with a power-law emissivity function:
\begin{equation}
\label{eq_eps}
\dustem=\epsilon(\lambda_0) \left( \frac{\lambda}{\lambda_0}
\right )^{-\beta}
\end{equation}
 where  $\epsilon(\lambda_0)$ is the emissivity at the reference wavelength $\lambda_0$,
and $\beta$ is the emissivity spectral index, usually taken as
$\beta=2$ (the so-called quadratic law). The quality of the available data
generally did not allow checking for variations in $\beta$, which was
therefore assumed to be constant.

The emission originates in dust radiating at thermal
equilibrium (at the temperature $T_d$) with the surrounding
radiation field, and as a consequence is related to the FIR and submm optical properties of the dust. Considering a constant gas/dust ratio $N_H/M_d$ (with $M_d$ the mass column density),
and assuming a single dust temperature along the line
of sight (LOS), the specific intensity can be written as
\begin{equation}
I_{\nu}(\lambda)=\kappa(\lambda_0) \left( \frac{\lambda}{\lambda_0}
\right )^{-\beta}B_{\nu}(\lambda,T_d)\times M_d,
\end{equation}
with
\begin{equation}
\kappa(\lambda_0)=\epsilon(\lambda_0) \frac{N_H}{M_d}
\end{equation}
where $\kappa(\lambda_0)$ is the dust mass opacity at the reference wavelength
$\lambda_0$.

In solid state physics, within the framework of classical
models, a constant and temperature-independent emissivity
spectral index value $\beta=2$ is justified by the asymptotic behavior
of the absorption cross-section of solids at long wavelengths,
far from the absorption resonances, such as the silicate
absorption features at 10-20 $\mic$. However, it was recognized early that the quadratic law may not hold for
amorphous materials in the FIR/submm, since such materials are likely
to exhibit low-energy transitions that happen in the FIR/submm
\citep[see][]{Andriesse74}. Although big grains
(BGs) are known to be amorphous in nature in the diffuse
ISM \citep{Kemper04}, no need was identified to explicitly
incorporate details about their internal structure in astrophysical
models due to the lack of submm data. However, recent observations
have proven that the actual FIR/submm dust SED is significantly more complicated than described by Eq. \ref{eq_eps}.

 Several recent studies have shown evidence of variations in the
dust emissivity law with wavelength. That the overall
Galactic SED is flatter than predicted by the quadratic law was
clearly shown in the FIRAS data by \citet{Reach95}. \citet{Paradis09}
studied the spectral dependence of dust emission in a collection of
molecular clouds and their atomic surroundings over the spectral range
covered by the DIRBE, Archeops, and WMAP experiments and showed that
the observed emissivity law exhibits a break around 500 $\mic$, with a
steeper index ($\beta$=2.4) in the FIR and a flatter index ($\beta$=1.5)
in the submm.  More recently a similar trend has been apparent
in the Planck data of the Taurus molecular cloud \citep{Abergel11}.
Similar submm excess emission has also been observed towards external
low-metallicity galaxies \citet{Galliano05}, as well as towards the Small
Magellanic Cloud in the Planck data \citep{Bernard11}. The comparison
between the spectral indices derived for the Galactic solar
neighborhood ($\beta=1.8$), the LMC ($\beta=1.5$), and the SMC
($\beta=1.2$) in the Planck data by \citet{Planck11,Bernard11} indicates a general flattening that could be linked to
the varying metallicity among these three environments. It therefore
appears that the dust SED is usually flatter than predicted by the
quadratic law and that the associated emissivity index $\beta$ varies
with wavelength and from object to object, possibly tracing
evolution of the dust properties with metallicity.

In addition to the above, the
spectral index of the dust emission as derived from a modified black-body fit to
the data follows a systematic trend with dust temperature, the
emissivity spectrum appearing flatter for regions with warmer dust
temperatures and steeper for colder regions.
\begin{itemize}

\item{\citet{Dupac03} show variations in $\beta$ in the FIR in various
regions of the ISM, using the PRONAOS balloon-borne
experiment data. They found variations in the index in the
range 2.4 to 0.8 for $\rm T_d$ between 11 and 80 K.}
\item{\citet{Desert08} highlighted variations in $\beta$ in the submm towards
compact sources in Archeops balloon-borne
experiment data. They found variations in the index in the range 4
to 1 for temperatures between 7 and 27 K.}
\item{\citet{Veneziani10} show variations in $\beta$ between 5 and 1 in
the temperature range 7-20 K by analyzing 8 high Galactic latitudes
clouds using BOOMERanG observations, combined with the IRAS, DIRBE, and WMAP
data.}
\item{\citet{Paradis10} found evidence of $\rm T_d$-$\beta$ variations
using the combination of the new Herschel and IRAS data, in the inner
regions of the Galactic plane. The $\beta$ range is 2.6-1.8 for
temperatures between 14 and 23 K.}

\end{itemize}
The origin of the flatness of the dust emission spectrum
observed with FIRAS has been explored by \citet{Reach95} and
\citet{Finkbeiner99}. The latter group propose a phenomenological description of
the FIRAS data that invokes the coexistence of a warm ($T_d
\simeq16\,K$) and a very cold dust ($T_d \simeq 9\,K$) component
(hereafter FDS model).
\citet{Reach95} determined that the FIR and the submm emission
were tightly correlated and therefore did not favor the very cold
dust component.

Moreover, the two component description does not reproduce the
observed variations in the emissivity spectral index with dust
temperature. Indeed the spectral index associated to each of the
components is set to a fixed value in the FDS model. For instance, in
their best-fit model, $\beta$ for the cold silicate-like component is
set to 1.7,  which does not allow SEDs to be produced with a larger
apparent emissivity index in the submm/mm domain (see Section \ref{sec_beta_predictions}).

The two-level systems (\TLS) model \citep[][hereafter Paper I]{Meny07} has been developed to describe the
FIR to mm continuum interstellar dust emission, taking the effect of the disordered internal structure of the amorphous dust
grains into account. It is based on the solid-state physics model developed to interpret
specific properties of the amorphous solids identified in laboratory data.
As a consequence, it is expected to apply with a high degree of
universality, and not to be sensitive to the exact chemical nature of
the dust. The disordered charge distribution (DCD) part of the model describes the interaction between
the electromagnetic wave and acoustic oscillations in the disordered
charge of the amorphous material \citep{Vinogradov60,Schlomann64}. This
charge-disorder is observed on nanometer scale and is described here by a single
charge correlation length. The TLS part of the model describes the interaction of
the electromagnetic wave with a simple distribution of asymmetric
double-well potential (ADWP), which results from disorder on atomic scale:
a ADWP can be associated with two close configurations
of atoms or group of atoms in the disordered structure
\citep{Phillips72, Phillips87, Anderson72}, where the corresponding atoms
have two equilibrium positions and can transit from one to the other.
Both DCD and TLS phenomena have been first applied by \citet{Bosch77}
to explain the observed temperature
dependence of the absorption of some silica-based glasses in the FIR/mm.

In section 2, we present the data we consider for this analysis. In
section 3, we briefly describe the model in terms
of the disordered structure, as well as the main parameters of the
model. In section 4, we explain the methods we used to compare the
model to astrophysical data and derive standard parameters for
the Milky Way (MW).  In section 5, we present model predictions for the
standard parameters derived above, such as the evolution of the dust
emission with temperature, the emissivity spectral index as a function
of wavelength and temperature, and the predicted emissivity in
Herschel and Planck photometric bands.  Sections 6 and 7 are devoted to
the discussion and conclusions, respectively.

\section{Astronomical data}

In the following we describe the dataset we used in this study.

\subsection{FIR/mm SED of the diffuse Milky Way emission}

\begin{figure*}
\begin{center}
\includegraphics[width=14cm]{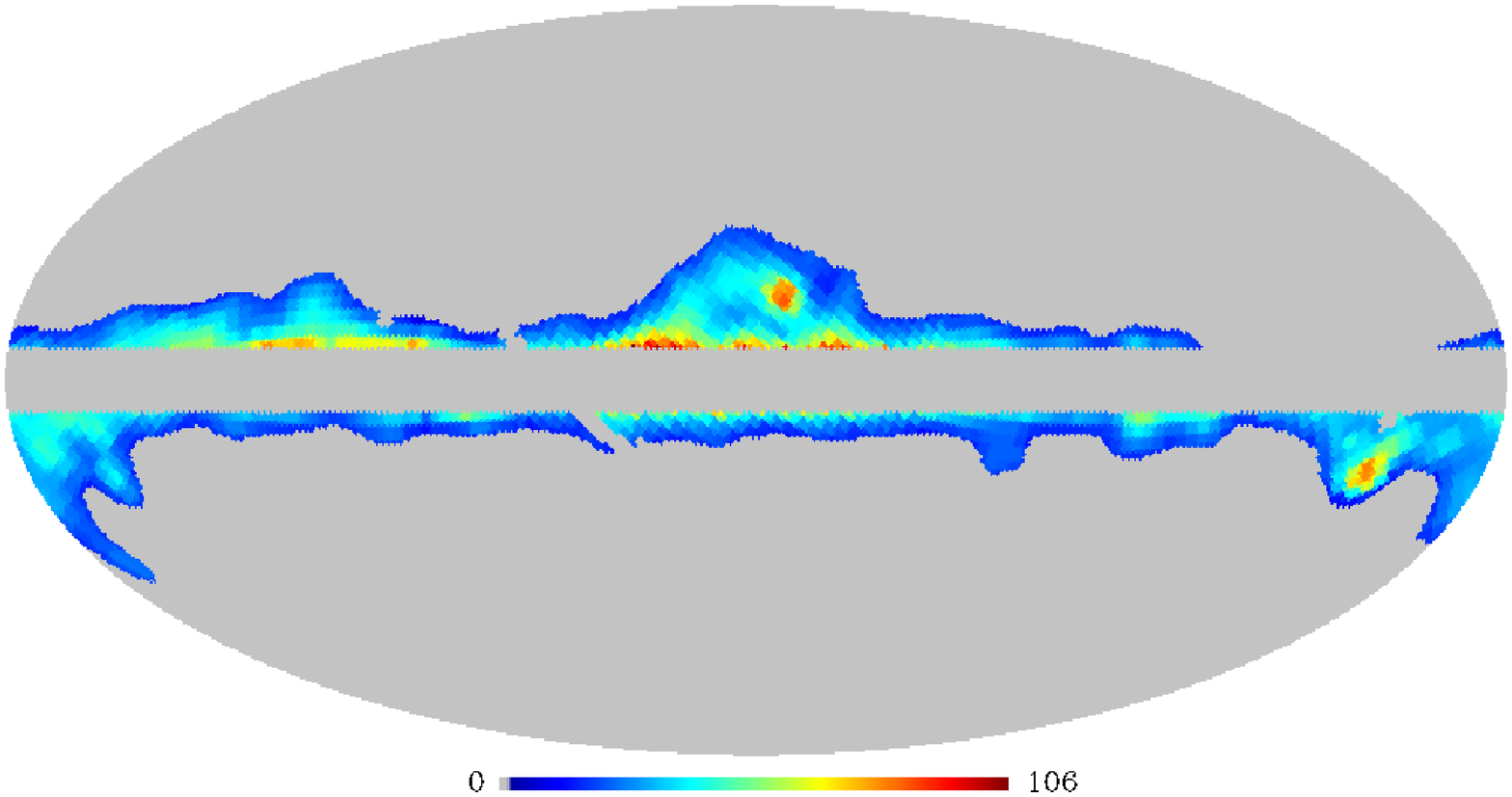}
\end{center} 
\caption{Map of the FIRAS 240 $\mic$ data used in this work
  ($|b|<6\degr$ and $\rm I_{240}<18$ MJy/sr). Units
  are MJy/sr. \label{fig_firas_240}}
\end{figure*}

The shape of the dust SED in the diffuse medium is an
important constraint on the physics of interstellar dust. In
principle, we would like to compare the new model with the SED of the
most nearby and the most diffuse ISM, in order to avoid
including emission from dense regions where dust properties may differ
from those in the diffuse solar neighborhood. Also, it is important to
analyze regions with minimum temperature mixing along the
LOS, which may make
the interpretations difficult. However, waiting for the Planck data to
become publicly available, the data presently available prevent us from
deriving such an SED up to millimeter wavelengths, because the low dust
emissivity in this range and the low dust column density available
combine into a poor signal to noise ratio (S/N). For instance, the solar neighborhood SED
presented by \citet{Boulanger96} did not allow measuring dust
emission in the solar neighborhood for $\lambda>1.1$ mm. Instead, we
used the emission from the whole MW, but exclude the
central parts of the Galactic disk ($|b|<6\degr$) in order to avoid
regions with extensive dust property mixing along the LOS,
as well as regions with $\rm I_{240}<18$ MJy/sr, to keep a reasonable
S/N. Figure \ref{fig_firas_240} shows the region of the sky
  seen by FIRAS at 240 $\mic$ that we
considered in this work, which accounts for 13.7$\%$ of the entire sky.

We constructed an average MW SED from 100 $\mic$ to 13 mm (23 GHz) by
merging data from the FIRAS instrument onboard the COBE satellite
and the WMAP data. We used the
full range of FIRAS wavelengths (0.1-10 mm or 3000-30 GHz) and the 5
WMAP bands at 13 mm (23 GHz), 9.1 mm (33 GHz), 7.3 mm (41 GHz), 4.9 mm
(61 GHz) and 3.2 mm (94 GHz). The angular resolution of the FIRAS data
is assumed to be $7\degr$. The angular resolution of the WMAP data is
taken from the WMAP explanatory supplement to range from 13$^{\prime}$
to 53$^{\prime}$ with increasing wavelengths.

For practical purposes, the FIRAS data were first reprojected into the
all-sky pixelisation scheme HEALPix (Hierarchical Equal Area
isoLatitude Pixelization) so that it can easily be combined with the
WMAP data. This was done using the method currently used for all
Planck ancillary data and described in \citet{Paradis11}.  The FIRAS data used here are available in the HEALPix scheme on
the WMAP LAMBDA (Legacy Archive for Microwave Background Data)
website\footnote{http://lambda.gsfc.nasa.gov/}. The WMAP data were
degraded to the FIRAS angular resolution ($7\degr$) through the
convolution with a Gaussian kernel of the appropriate FWHM using the
standard convolution tools in HEALPix, before extracting the SED. An offset subtraction
was done by removing an average SED computed over regions of the
sky with low emission set as $\rm I_{240} < 3$ MJy/sr, corresponding
to high Galactic latitudes. The method used removes any arbitrary additive offset in the data
that could be set by calibration uncertainty or cosmological sources
such as the cosmic infrared background (CIB). Uncertainties on the SED are computed as
the quadratic sum of three different uncertainties for each
instrument: calibration, white noise per pixel, and background. We assumed the official calibration and white noise uncertainties
given in the explanatory supplements. Calibration uncertainties are
$\leq$10$\%$ for FIRAS, depending on the frequency range, and less than
1$\%$ for WMAP data \citep{Jarosik11}. The background uncertainties on
the FIRAS and WMAP data have been derived by computing the standard
deviation over fifty circular regions with a 2$\degr$ radius, taken
randomly in the high latitude sky. This uncertainty largely
predominates over the other ones.

The obtained SED is shown in Figure \ref{fig_tls}, and the SED values
and uncertainties are given in Table\,\ref{tab_spec_firas}. The
corresponding average HI column density is 2$\times 10^{21}$ H/cm$^2$. It can be
seen from Figure \ref{fig_tls} that the emission peaks near
$\lambda=160\mic$, in agreement with the FIRAS SED derived at high
latitudes by \citet{Boulanger96}, who found an average dust
temperature of 17.5 K assuming a constant spectral index set to 2.
The recent analysis of the Planck data shows
that a more appropriate value for the FIR spectral index in the solar
neighborhood is $\beta$=1.8 \citep{Bernard11}, and
using this value over nine fields in the galactic halo leads to a similar
temperature value of 17.9 K \citep{Miville11} and 17.7 K for regions
with $|b|>10\degr$. 
Above $\lambda \simeq 5$ mm, the ISM emission starts increasing
with wavelengths.  This contribution does not originate in thermal
dust emission, but is most probably from a combination of
synchrotron or free-free emission and/or to anomalous microwave
emission, which is often attributed to spinning dust \citep[see for
instance][]{Finkbeiner04,Dickinson11,Marshall11}

\begin{table*}[!ht]
\begin{center}
\begin{tabularx}{\textwidth}{lXXXXXXXXXXXXX}
\hline
\hline
$\lambda$ ($\mu$m)& 104.0 & 106.5 & 109.2 & 111.9 & 114.9 & 117.9 &
121.2 & 124.6 & 128.2 & 132.1 & 136.1 & 140.5 & 145.1 \\
Brightness (MJy/sr)& 17.08 & 24.79 & 26.28 & 30.92 & 30.73 & 33.27 & 32.93
& 34.55 & 35.72 & 36.40 & 37.87 & 38.86 & 39.34   \\
1-$\sigma$ (MJy/sr)& 49.38 & 15.95 & 11.24 & 11.66 & 7.22 & 7.66 & 6.48
& 6.49 & 4.58 & 4.37 & 3.13 & 2.75 & 2.28 \\
\hline
\end{tabularx}
\begin{tabularx}{\textwidth}{lXXXXXXXXXXXXX}
\hline
$\lambda$ ($\mu$m) & 150.0 & 155.3 & 161.0 & 167.1 & 173.6 &
180.8 & 188.5 & 196.9 & 206.1 & 216.2 & 227.3 & 239.7 & 253.5 \\
Brightness (MJy/sr) & 40.15 & 39.90 & 41.82 & 40.76 & 40.86 & 40.29
& 39.33 & 38.13 & 36.66 & 34.68 & 32.51 & 30.18 & 27.57 \\
1-$\sigma$ (MJy/sr)&  2.75 & 2.20 & 1.39 & 1.29 & 0.99 & 1.50
& 1.16 & 0.91 & 0.53 & 0.44 & 0.35 & 0.28 & 0.29 \\
\hline
\end{tabularx}
\begin{tabularx}{\textwidth}{lXXXXXXXXXXXXX}
\hline
$\lambda$ ($\mu$m) & 268.9 & 286.4 & 306.3 & 329.1 & 355.7 & 386.9 & 424.1 & 469.2 &
525.1 & 596.0 & 698.1 & 816.8 & 1002.4  \\
Brightness (MJy/sr) & 24.92 & 22.12 & 19.35 &16.62 & 14.01 & 11.53 & 9.37 & 7.22 & 5.21 &
3.66 & 2.46 & 1.49 & 0.79 \\
1-$\sigma$ (MJy/sr) & 0.25 & 0.35 & 0.22 & 0.24 & 0.24 & 0.24 & 0.29 & 1.01 & 0.33 & 0.17 & 0.12 & 0.11 & 0.09 \\
\hline
\end{tabularx}
\begin{tabularx}{\textwidth}{lXXXXXXXX}
\hline
$\lambda$ ($\mu$m) & 1297.2 & 1837.7 & 3150.3 &  3200.0 & 4900.0 & 7300.0 & 9100.0 & 13000.0  \\
Brightness (MJy/sr) &0.35 & 0.13 & -1.6$\times$10$^{-2}$ & 1.9$\times$10$^{-2}$ & 9.2$\times$10$^{-3}$ & 9.8$\times$10$^{-3}$ & 1.1$\times$10$^{-2}$ & 1.5$\times$10$^{-2}$ \\
1-$\sigma$ (MJy/sr) & 0.10 & 0.10 & 0.12 & 1.1$\times$10$^{-3}$ & 5.6$\times$10$^{-4}$ & 2.7$\times$10$^{-4}$ & 2.3$\times$10$^{-4}$ & 1.8$\times$10$^{-4}$ \\ 
\hline
\end{tabularx}
\end{center}
\caption{FIRAS/WMAP SED of the diffuse MW emission. \label{tab_spec_firas}}
\end{table*}

\subsection{FIR/submm SEDs of Galactic compact sources}
\label{sec_cs}
Before the Planck data become publicly available, the FIRAS
data are the only data allowing the dust
equilibrium temperature and the FIR-submm spectral index of dust
emission to be measures simultaneously over large regions of the sky. This is mainly because this
requires both FIR data (which measure temperature best) and submm data
in the Rayleigh-Jeans regime, which constrains the submm
emissivity index best. The FIRAS data accurately sample the peak of
the BG emission. A good sampling of data in the range 100- 500 $\mic$
is crucial for constraining the dust temperature, as well as the
correlation length, which is a free parameter of the \TLS model
(see Section \ref{sec_dcd}). Unfortunately, at the angular resolution
of the FIRAS instrument, the sky distribution provides very little
handle on dust temperature variations. At the FIRAS resolution, the
dust temperature varies only from 16 to 21 K \citep{Reach95}, assuming
a constant spectral index equal to 2. Therefore, testing the
temperature dependence of the dust spectral index requires additional
information in the submm over wide dust temperature range.

Balloon experiments have been able to record FIR/submm data at a
significantly better angular resolution than for the COBE and
FIRAS data.  Both the PRONAOS and the Archeops data analysis
\citep{Dupac01,Desert08} show a decrease in the spectral
index with temperature. The former study used data of the diffuse
emission mapped over limited regions going from cirrus clouds to dense
regions. The latter study was carried using observations of
compact sources in the MW. The variations sample the range from
$\beta \simeq 4$ for compact sources with dust temperatures as low as
7 K down to $\beta \simeq 1$ for the highest temperatures of about 27
K. The interpretation of the observed variations in the
emissivity index with temperature is still being debated.  \citet{Shetty09}
argue that the observed anticorrelation could result entirely from
noise in the data or from temperature mixing along the LOS.  However, the recent analysis of the Herschel
data in the Galactic plane by \citet{Paradis10} demonstrated that the
observed variations could not be entirely due to data
noise. Similarly, the analysis of the Planck data towards cold cores
showed that the expected LOS temperature mixing could not explain the
whole observed trend.  It is therefore likely that the observed
steepening of the emissivity index with decreasing temperature reveals
changes in the optical properties of astronomical dust with
temperature. Laboratory experiments \citep{Boudet05,
Coupeaud11} have also shown evidence of a similar
behavior for analogs of the amorphous material likely to compose
astrophysical dust grains, which tend to favor actual variations in
the dust properties with temperature.

Here, we do not intend to fit any given $\rm T_d$-$\beta$
relationship. Instead, we minimize against the overall Archeops
compact source spectra, which do not impose following a
predetermined $\rm T_d$-$\beta$ relationship. Archeops observed about 30$\%$
of the sky at a 12$^{\prime}$ angular resolution, in four photometric
bands: 550, 850, 1382 and 2098 $\mic$ \citep{Benoit03}. 
The Archeops compact sources catalog \citep{Desert08} contains 302 SEDs, which have been fitted using a modified
  black body. Only 153 fits have given a satisfactory
  $\chi^2$ (using a $\chi^2$ goodness-of-fit criteria at the
  2-$\sigma$ level) when using free $\rm T_d$
  and $\beta$ \citep{Desert08}. We therefore restricted our
  analysis to these 153 spectra.  Since the sampling of the compact sources
  spectra is not evenly distributed in dust temperatures, we averaged
them in ten bins of temperature, between 7.3 and 25.7 K. In this way,
we ensure a similar weight for the ten spectra. The summary of
the binning parameters, including the bin's central temperature, the
average fluxes, the
number of sources, as well as their Archeops names, are provided in Table \ref{tab_arch}.

The uncertainty on the data is the quadratic combination of the
relative and the absolute uncertainties (15, 17, 21, and 14 $\%$ from 550
$\mic$ to 2.1 mm).

\begin{table*} 
\begin{center}
\resizebox{\textwidth}{!}{%
\begin{tabular}{l c c c c c c c c}
 \hline
\hline
 Bin & $T_{fit} $ &$F_{100}$ & $F_{550}$ & $F_{850}$ & $F_{1382}$ & $F_{2098}$ &
 nb. compact sources/bin & Arch. name\\
  & (K) & (Jy) & (Jy) & (Jy) & (Jy) & (Jy) & & \\
\hline
1&  8.2 & 25.1$\pm$3.4 & 434.8$\pm$66.0  &  86.5$\pm$15.0  & 15.1$\pm$3.2 &
3.1$\pm$0.6 &  29 &  15, 41, 48, 63, 64, 66, 73, 110,124, 132, \\
& & & & & & & & 143, 151, 158,
188, 191, 200, 222, 231, 232,
239, \\
& & & & & & & & 245, 247, 256, 259, 266, 272, 289, 291, 297\\
2&  10.1 & 34.2$\pm$4.6 & 315.8$\pm$47.8  &  65.6$\pm$11.3 & 15.3$\pm$3.2  &
3.0$\pm$0.5 & 47 & 1, 45, 49, 51, 54, 69, 79, 87, 90, 99, \\
& & & & & & & & 105, 106, 111, 137, 138,
139, 141, 145, 146, 150, \\
& & & & & & & & 169, 171, 174, 175, 176, 180, 181, 186, 
192, 197, \\
& & & & & & & & 203, 204, 210, 227, 228, 
240, 249, 252, 255, 267, \\
& & & & & & & & 270, 274, 276, 284, 295, 296, 298,  \\
3& 11.9 & 99.6$\pm$13.4 & 380.9$\pm$58.5  & 81.9$\pm$14.4  & 21.9$\pm$4.6  &
4.7$\pm$0.8  &  20 & 17, 37, 47, 58, 62, 65, 84, 93, 116, 127, \\
& & & & & & & & 162, 168, 196,
206, 207, 211, 220, 224, 242, 253\\
4&  13.8 & 368.1$\pm$49.7 & 416.8$\pm$63.3  & 106.3$\pm$18.3  & 22.0$\pm$4.7
& 5.5$\pm$0.9  &  21 & 38, 59, 61, 76, 89, 97, 98, 109, 130, 155, \\
& & & & & & & & 157,
160, 166, 173, 185, 189, 216, 229, 251, 260, \\
& & & & & & & & 292\\
5&  15.6 & 700.8$\pm$94.6  & 410.1$\pm$62.8  & 96.2$\pm$17.3 & 23.2$\pm$4.9
& 5.9$\pm$1.0 &  12 & 21, 35, 95, 102, 113, 118, 163, 164, 235, 237,
\\
& & & & & & & & 275, 301,  \\
6& 17.4 & 1960.7$\pm$264.7  & 573.3$\pm$88.0 & 145.6$\pm$25.3 & 29.8$\pm$6.3
& 8.1$\pm$1.3 &  9  & 11, 14, 112, 133, 152, 269, 277, 290, 300 \\
7&  19.3 & 3311.4$\pm$447.0  & 579.6$\pm$89.1  & 133.0$\pm$23.3 &
28.6$\pm$6.1  & 5.9$\pm$1.2 &  5  & 78, 92, 100, 226, 246 \\
8&  21.1 & 1877.5$\pm$253.5  & 367.3$\pm$61.2   & 95.5$\pm$17.9 &
29.9$\pm$6.5  & 5.6$\pm$1.3 &  4  & 5, 117, 179, 268\\
9&  23.0 &  2720.0$\pm$367.2  & 730.0$\pm$127.3 & 222.0$\pm$40.6  &
54.2$\pm$12.0  &  19.3$\pm$ 3.3 & 1  & 234\\
10& 24.8 & 991.0$\pm$133.8  & 257.5$\pm$44.9  & 66.5$\pm$13.5 & 23.9$\pm$5.3
& 8.0$\pm$1.4   & 2  & 57, 114\\
\hline
\end{tabular}}
\end{center}
\caption{Description of the Archeops spectra bins used in this study.  \label{tab_arch}}
\end{table*}

\section{The \TLS model}

In the following, we first remind the reader of the relation between
dust emission and absorption and then briefly summarize the different processes of the \TLS
model. The full theoretical description of the model is given in Paper
I. 

\subsection{Relation between dust emission and absorption}

 The Mie theory is based on the exact resolution of the
Maxwell's equation for the absorption and scattering of a
plane electromagnetic wave by a homogeneous spherical particle
whose material is characterized by its complex dielectric
constant $\epsilon(\omega) = \epsilon^{\prime}(\omega) + i
\epsilon^{\prime \prime}(\omega)$, with $\omega=2\pi c/\lambda$. Solid
state physical models often analyze the optical properties by
the use of the complex susceptibility $\chi_0(\omega) =
\chi^{\prime}(\omega)+i \chi^{\prime \prime}(\omega)$,
which in the case of isotropic dielectric materials is related to the
complex dielectric constant by the Clausius-Mossotti equation: 
\begin{equation}
\epsilon(\omega)=1+\frac{4\pi \chi_0(\omega)}{1-\frac{4\pi \chi_0(\omega)}{3}}.
\end{equation}
For spherical grains of radius $a$ and density $\rho$, the Mie theory
gives the following expression for the dust mass opacity:
\begin{equation}
\kappa(\omega)=\frac{9\omega}{\rho c}Im\left( \frac{\epsilon(\omega)-1}{\epsilon(\omega)+2} \right ).
\end{equation}
The optical properties of bulk materials are also often characterized
by the absorption coefficient $\alpha$, which is defined as
the optical depth $\tau$ in the bulk material per unit length. The
absorption coefficient $\alpha$ is related to the complex dielectric
constant $\epsilon$ by
\begin{equation}
\alpha(\omega)=2Im\left( \frac{\omega \sqrt{\epsilon}}{c}\right ).
\end{equation}
In general, the complex dielectric constant is wavelength-dependent,
and as a consequence $\alpha(\omega)$ is not proportional
to $\kappa(\omega)$, the ratio $\kappa/\alpha$ is
wavelength-dependent.

However, for dielectric dust materials in the FIR/submm
range, the value of the imaginary part $\epsilon^{\prime \prime}$ of the complex dielectric
constant $\epsilon(\omega)$ is negligible with respect to the real
part $\epsilon^{\prime}(\omega)$ (details can be found
in Paper I). This condition $\epsilon^{\prime \prime}(\omega) \ll
\epsilon^{\prime}(\omega)$ leads to
\begin{equation}
\alpha(\omega) \approx \frac{\omega}{c} \frac{\epsilon^{\prime \prime}}{\sqrt{\epsilon^{\prime}}},
\end{equation}
and therefore:
\begin{equation}
\frac{\kappa(\omega)}{\alpha(\omega)}=\frac{9
  \sqrt{\epsilon^{\prime}}}{\rho (\epsilon^{\prime}+2)^2}.
\end{equation}
The real and imaginary parts of the complex dielectric constant
are coupled through the Kramers-Kronig relations. It
implies that $\epsilon^{\prime \prime}(\omega) \ll
\epsilon^{\prime}(\omega)$ and that the real
part of $\epsilon(\omega)$ can be considered constant
($d\epsilon^{\prime}/d\omega=0$). As a consequence, the ratio
$\kappa(\omega)/\alpha(\omega)$ is almost constant in the
FIR/submm range, and the slopes of $\kappa(\omega)$ and $\alpha(\omega)$ are
the same.

\subsection{A disordered charge distribution}
\label{sec_dcd}
The disordered charge distribution (DCD) description of an
amorphous material shows that in the FIR/mm wavelength range, the
absorption coefficient due to acoustic lattice oscillations is temperature-independent and can be written:
\begin{equation}\label{eq:S64}
\alpha_{DCD}
=\frac{(\epsilon+2)^2}{9} \frac{1}{3\ \vtrans^3}\langle
\frac{q^2}{m}\rangle \frac{\omega^2}{c\sqrt{\epsilon}}
\left[1-\left(1+\frac{\omega^2}{\omega_0^2}\right)^{-2}\right].
\end{equation}
Here, $\omega_0=v_t/l_c$ with $\rm v_t$ the transverse sound
velocity in the material and $\rm l_c$ the charge correlation length,
c is the speed of light, $\rm <q^2/m>$ the mean squared
charge deviation per atomic mass, and $\epsilon$ the dielectric constant. It must be pointed out that the DCD
process presents two asymptotic behaviors for the absorption and
spectral index, on both sides of $\rm \omega_0$.The spectral index varies between
2 in the short wavelength range ($\rm \omega>\omega_0$) and 4 in the
long wavelength range ($\rm \omega<\omega_0$). $\epsilon$, $\rm
v_t$, and $\rm <q^2/m>$ is likely to vary with material
composition and with disorder at second order. In the absence of a
clear identification of the material composing astronomical grains,
we adopt for modeling purpose the numerical values used by
\citet{Bosch77}.
The important effect for our study is the wavelength profile of
the DCD absorption. As a consequence only the numerical value of
$\rm \omega_0$ is relevant. The value of $\rm l_c$ will be derived
from $\omega_0=v_t/l_c$ taken the reference sound
velocity equal to 3$\times10^5$ cm.s$^{-1}$ into account. 

\subsection{A two-level system}
\label{sec_tls}
Originally, the TLS description is based on the assumption that
the amorphous characteristics of the solid enables close
configurations of atoms or groups of atoms, which can be described by
an ADWP. The TLS model describes the interaction between
electromagnetic waves and the split ground states (which represent
the two-level systems) of these ADWP that are optically active;
i.e. they are characterized by a non-zero dipole
moment. Phenomenologically, because of the amorphous nature of
the material, each ADWP is local and specific, but in a macroscopic
analysis, the disorder in the amorphous material can be studied
in terms of an ordered distribution of TLS with a simple density of
states. Three interaction mechanisms, which all
depend on temperature, can occur in such a population of TLS sites: a resonant absorption, and two relaxation processes identified as ``tunneling relaxation'' and ``hopping
relaxation''. As opposed to resonant phenomena, the
electromagnetic field does not cause the transition between the two
energy levels for relaxation mechanisms. It means that all relaxation
processes, regardless of their splitting energy $\hbar \omega$, can
participate in the absorption of the electromagnetic wave of a given
frequence $\omega_0$, unlike the resonant effect where only the TLS
characterized by the energy $\hbar \omega_0$ can absorb.
The various effects can be described as follows:
\begin{itemize}
\item{Resonant absorption \citep{Hubbard03}:\\
The resonant absorption of a photon of energy $\hbar\omega$ concerns
only those TLS characterized by the splitting energy $\hbar
\omega$. Such a mechanism gives an absorption that
decreases with the increasing temperature and vanishes when the
temperature is high enough that the two-levels become equally
populated.  The resonant absorption can be written as:
\begin{equation}
 \alpha_{res} = \frac{4\pi^2}{3c\sqrt{\epsilon}}\frac{(\epsilon+2)^2}{9}\;
\omega G(\omega)\;\tanh(\hbar\omega/2 k\Tdust),
\end{equation}
where k is the Boltzmann constant, $\epsilon$ the dielectric
constant, and $G(\omega)$ the optical density of state (ODOS).
For simplicity, and to compare our results with a reference, we follow
\citet{Bosch77} and consider a
constant ODOS, i.e. $G(\omega)=G_0$, with $G_0$ equal to
1.4$\times 10^{-3}$ (CGS units).}\\

\item{Tunneling relaxation:\\
In terms of absorption, this process occurs when the phonon energy is
too low to allow direct transitions above the barrier height between
the two levels. In this case, additional tunneling is required to
perform the transition. The tunneling absorption increases with
temperature. We derive the following accurate and relatively
simplistic expression for the phonon-assisted tunneling absorption,
more convenient for computation than the formula obtained by
\citet{Fitzgerald01a,Fitzgerald01b}:
 \begin{eqnarray}\label{eq:FC01}
&&\alpha_{phon} = \frac{G_0}{3c\sqrt{\epsilon}} \frac{(\epsilon+2)^2}{9}
  \omega\ F_2(\omega,T_d), \\
&& \label{eq:f2wT} F_2(\omega,T_d) =  \\
&&=\frac{1}{2kT_d} \nonumber \int^{\infty}_{0}
\int^{\infty}_{\tau_1}\sqrt{1-\frac{\tau_1}{\tau}}\
\sech^2(\frac{E}{2kT_d}) \frac{\omega d\tau dE}{1+\omega^2\tau^2},
\label{eq:f2i}
\end{eqnarray}
where $\tau_1$ is the relaxation time defined by $ \tau_1= a\
E^{-3}\tanh(E/2k\Tdust).$ A reasonable value for $a$ is $4.2\times
10^{-56}$ erg$^3$ s \citep[in the case of soda-silica glass;][]{Bosch77}}.\\

\item{Hopping relaxation:\\
Hopping relaxation occurs when the phonon energy allows a direct jump
above the barrier. It gives an absorption that increases with temperature: 
\begin{eqnarray}\label{eq:B78H}
\alpha_{hop} &=& \frac{8\pi}{3c\sqrt{\epsilon}}
\frac{(\epsilon+2)^2}{9} G_0  (c_{\Delta} + lnT_d) \int^{\infty}_{0} dV P(V)
\frac{\omega^2\tau}{1+\omega^2\tau^2}
\end{eqnarray}
where $\tau$ is the relaxation time defined by $\tau = \tau_0
\exp(V/T_d)$. For amorphous silica, $\tau_0\simeq 1
\times 10^{-13}$s \citep{Bosch77}.
$c_{\Delta}$, an additional parameter of tunneling states is given by
\begin{equation}
c_{\Delta}=ln \frac{k_B}{\Delta_0^{min}}+ln4-1+\int_0^1 ln\,arctan(x)\,dx
\end{equation}
where $\Delta_0^{min}$ is the minimal energy splitting. Only a lower
limit of $c_{\Delta}$ with a value close to 5.8 has been determined from laboratory experiments. 
P(V) is the distribution of the TLS barrier height which can be
approximated by
\begin{equation}
P(V)=C_Vp_v(V)  \textrm{ with }
\end{equation}
\begin{equation}
C_V \cong \frac{1}{V_0 \sqrt{\pi}}, \textrm{ and }
\end{equation}
\begin{equation}\label{eq:p(V)}
p_v(V)= \left\{ \begin{array}{ll}
\exp(-(V-V_m)^2/V_0^2) &\textrm{ for }V>V_{min},\\
0 &\textrm{ for }V<V_{min},
\end{array} \right. \end{equation} 
such as $V_{min}=50$ K, $V_m =550$ K and $V_0 =410$ K. }
\end{itemize}

All physical parameters are set to the values from
\citet{Bosch77}, and are provided in Table \ref{tab_physical_param}.
The amplitude of the TLS processes with respect to the DCD part is
controled by a relative intensity parameter denoted $A$ (see Section \ref{sec_param}). The total absorption ($\alpha_{tot}$) deduced from the
\TLS model is the sum of all the processes quoted in this section:
\begin{equation}
\label{eq_tot}
\alpha_{tot}=\alpha_{DCD}+A\times(\alpha_{res}+\alpha_{phon}+\alpha_{hop}).
\end{equation}
\begin{figure}
\begin{center}
\includegraphics[width=8cm]{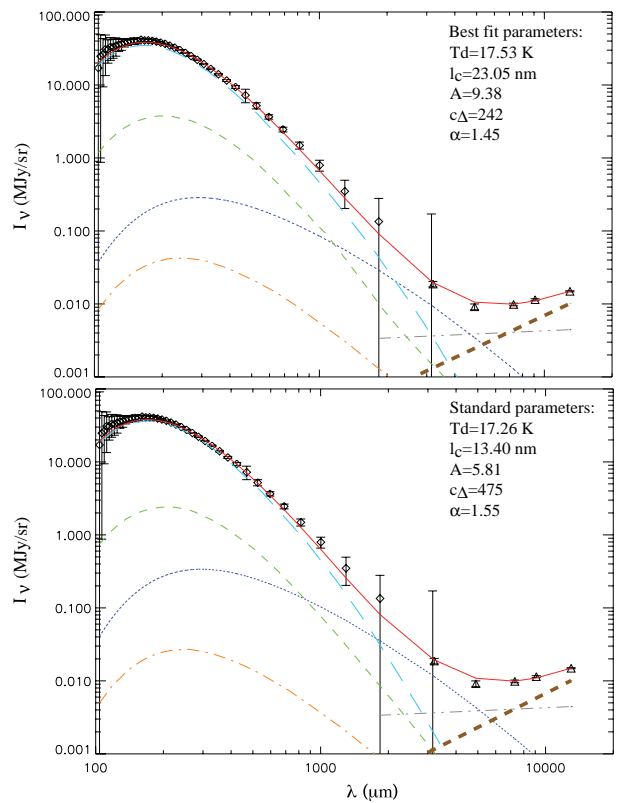}
\end{center} 
\caption{FIRAS/WMAP diffuse MW spectrum fitted using the \TLS model: total in red (solid), DCD process in light blue (long
dash), and TLS processes:
resonant absorption in green {\bf (dash)}, hopping relaxation in dark
blue (dot), and
tunneling relaxation in orange (dash-dot), with the best-fit parameters (top panel)
and standard parameters (bottom panel). The gray line (dash-dot-dot-dash) corresponds to
free-free emission, and the brown line (thick dash) to a $\lambda^{\alpha}$
power law.\label{fig_tls}}
\end{figure}

\subsection{Model parameters}
\label{sec_param}
The relevant parameter for the DCD absorption is the charge correlation
length $l_c$, which controls the wavelength where the inflection point
between the two asymptotic behaviors with $\beta=2$ and
$\beta=4$ occurs. The relative amplitude between the DCD and the
TLS effects is not strongly constrained by the model prior
to the comparison with astrophysical data. The factor that controls the amplitude of the DCD absorption depends on the
characteristics of the material, whereas the amplitude of the TLS
effects (resonant, tunneling, and partly hopping) is related to the TLS
density of states. Therefore, to simply characterize the relative importance of the TLS and the DCD processes, we use
a single free parameter of relative intensity denoted $A$. This means
that all TLS effects are multiplied by the same intensity $A$ (see
Eq. \ref{eq_tot}). The
respective contributions of the resonant and of the two relaxation
processes are set according to theory. The resonant effect and
tunneling relaxation correspond to a quantum tunneling effect. However,
the hopping relaxation is a more classical effect, since there is no
tunneling involved.
The characteristics of the potential wells to be taken into account in
computing the absorption in case of hopping or tunneling are different, depending on the
classical or quantum tunneling process. Therefore, we want to keep the
possibility of uncoupling these effects, by means of the $c_{\Delta}$
coefficient.
The dust temperature can be derived from the shape
of the dust SED in the FIR and also controls the amplitude of the TLS
relaxation effects.  It is therefore a crucial parameter of the model,
linking the FIR emission to the submm behavior of the optical
properties of dust.

Finally, we adopt the following four free parameters for the \TLS
model in this study: $T_d$, $l_c$, $A$, and $c_{\Delta}$. Note
that these parameters allow only to characterize the shape of the
emission. One additional parameter is required to set the overall
amplitude of dust emission. Classically this is done by assuming a
dust/gas ratio and some dust emissivity at a given reference
frequency. Also, the number of free parameters used here
is roughly the same as the one needed to achieve a description of the
emission shape with the FDS model
($\alpha_1$, $\alpha_2$, $T_2$, $q_1/q_2$, and $f_1$).
\begin{table*}[!t] 
\begin{center}
\begin{tabular}{l c c c c c c c c}
\hline
\hline
$v_t^*$ & $<q^2/m>^*$ & $\epsilon^{*\dag}$ & $G_0 ^\dag$ & $a ^\dag$ & $\tau_0 ^\dag$ & $V_0 ^\dag$ &
$V_m ^\dag$ & $V_{min}^\dag$ \\
(cm/s) & erg.cm.g$^{-1}$& & (CGS units) & (erg$^3$.s) & (s) & (K) &
(K) & (K) \\
\hline
3$\times10^5$ &  8748 & 2.6 & 1.4$\times10^{-3}$ &
4.2$\times10^{-56}$ & $10^{-13}$ & 410 & 550 & 50 \\
\hline
\end{tabular}
\caption[]{Physical parameters used in the \TLS
  model. \\
$^*$ defined in Section \ref{sec_dcd}.\\
$^\dag$ defined in Section \ref{sec_tls}.\\
\label{tab_physical_param}} 
\end{center}
\end{table*}

\section{Comparison with astrophysical data}
\subsection{Determination of the best-fit parameters}
The diffuse medium sampled by the FIRAS/WMAP SED and the compact
sources detected in the Archeops data could exhibit rather different
dust properties. In a first step, we therefore model the two datasets
separately, in order to illustrate possible differences between the two
environments. However, as explained in Section \ref{sec_cs}, only the
Archeops data sample the range of temperature needed to 
constrain the \TLS model. We also therefore perform a combined fit
where all data are used, in order to derive standard parameter values
allowing both diffuse medium and compact sources to be reproduced. In each
case, we proceed in the same way, i.e., we use the idl ``mpfit''
function to perform a $\chi^2$ minimization between the observational data and the \TLS model. Results of the minimization are
given in Table \ref{tab_bootstrap}.  The prediction of the model is
color-corrected to match the observation in the large photometric filters
of each instrument.

\subsubsection{Modeling of the FIRAS/WMAP MW spectrum}
\label{sec_firas}
The model has been compared with the FIR/mm SED of the diffuse MW
emission. For the longest wavelengths, dust emission is not the
main contributor, in particular in the WMAP bands. We take free-free emission into
account using the free-free template at 30 GHz
constructed by \citet{Dickinson03}, derived from the H$\alpha$ map,
and extrapolated in frequency using a power law for the
  specific intensity with $I_{\nu} \propto \lambda^{0.1}$. In addition to the free-free emission, we have
incorporated a power law emission that can describe synchrotron
emission, probably mixed with other types of emission such as
spinning dust. The slope of  this power law ($\alpha$) is left as
a free parameter, since contamination in this wavelength range is not
well known. The brightness level of the free-free combined with the power law
emission has been normalized to the 23 GHz data. Figure \ref{fig_tls}
(top panel) shows the result of the
modeling. From Table \ref{tab_bootstrap} we can see that the dust
temperature is robustly derived, and the value of 17.5 K is similar to
the results of \citet{Boulanger96}, who analyzed the emission in
the solar neighborhood. However the other parameters are not well
constrained using only the FIRAS/WMAP SED. The $A$ parameter controls the intensity of all TLS effects, whereas the
$c_{\Delta}$ parameter allows us to increase/decrease the emission
coming from the hopping relaxation. To reproduce only the mm flattening in
the spectrum deduced from the
FIRAS data, the model requires an increase in the emission in the
submm domain, which can be explained by an increase of either $A$ or
$c_{\Delta}$. These two
parameters are therefore highly degenerate in this fit
configuration. The 1-$\sigma$ uncertainty on the correlation length
indicates that using only the FIRAS/WMAP data is not sufficient for
constraining this parameter, which controls the slope of the emission
spectrum. Indeed, changing $l_c$ is equivalent to modifying the 
apparent emissivity spectral index $\beta$, and $\beta$ is known
to be naturally anticorrelated with dust temperature. Therefore, the
only way to constrain this parameter is to compare the model to
spectra at various temperatures, which means that the modeling of
the Archeops spectra is important to derive $l_c$ (see Section \ref{sec_arch}).
The specific intensity is best fitted by a power law of the form
  $I_{\nu} \propto \lambda^{\alpha}$ with $\alpha=1.45$. Variations in the synchrotron
index have been found in our Galaxy with values ranging from 0.6
in star-forming complexes to 1.2 in the interarm regions. Our value is
higher than the observed $\alpha$ values for synchrotron alone,
which could indicate that this emission is not only due to
synchrotron emission, but may also include a significant
contribution from spinning dust emission. Although spinning dust
emission has not been shown at high latitude, some studies
indicate that it is widespread in the Galaxy \citep{Miville05,Marshall11}.
\begin{figure*}
\begin{center}
\subfigure{\includegraphics[width=16cm]{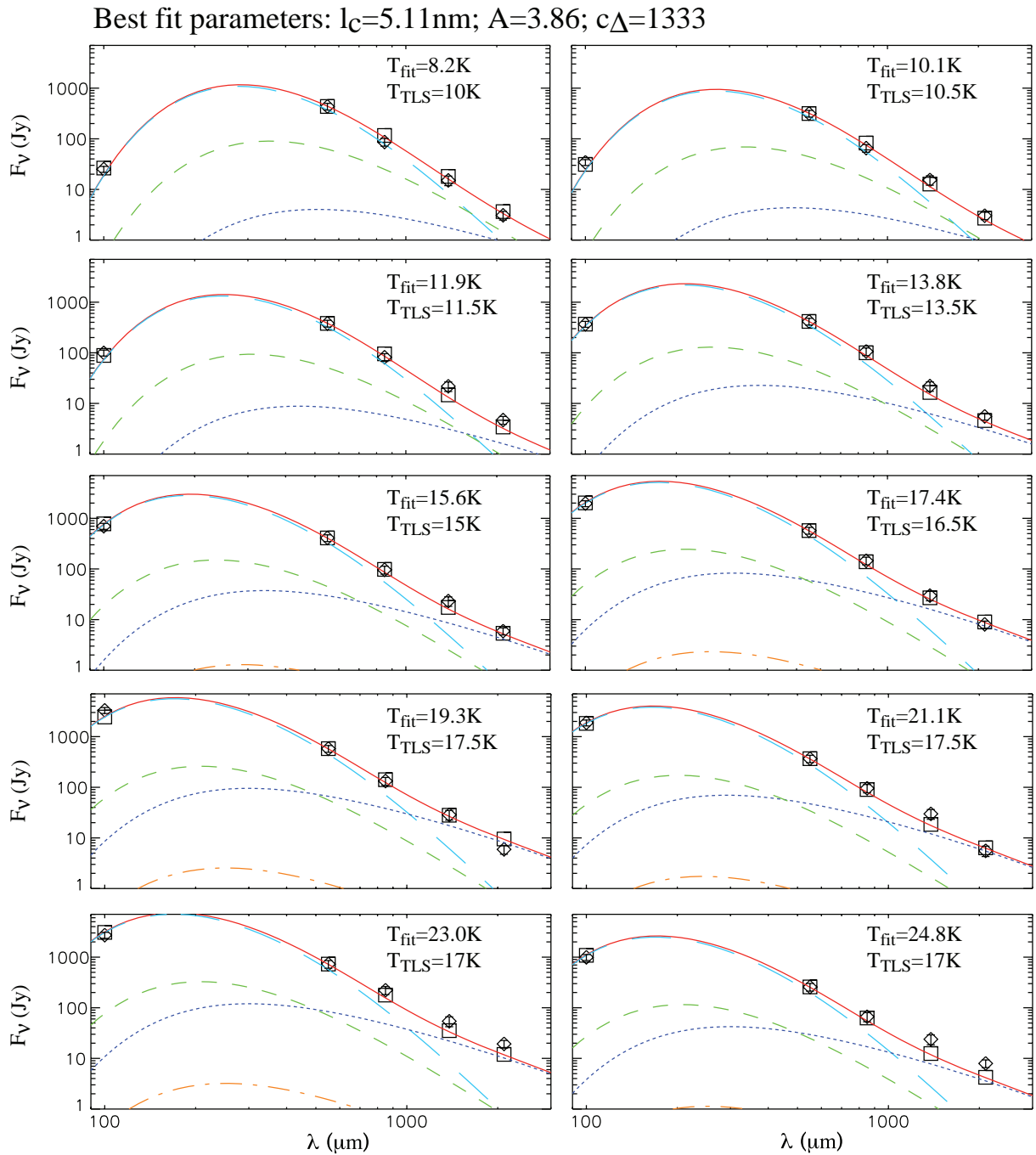}}
\caption{Spectra of Archeops compact sources (diamonds), binned in
temperature, fitted using the
\TLS model, with the best-fit parameters (this page) and the standard
parameters (next page). The contribution of the different processes is given
in the caption of Figure \ref{fig_tls}. Squares correspond to the model
after color correction. The average temperature per bin deduced from
a single modified black-body fit (from D\'esert et al.,2008), and the temperature
deduced from the \TLS model are given in the top right of each
panel. The description of the Archeops spectra is given in Section
2.2 and in Table 2. \label{fig_arch}}
\end{center} 
\end{figure*}

\addtocounter{figure}{-1}
\begin{figure*}
\addtocounter{subfigure}{1}
\begin{center}
\subfigure{\includegraphics[width=16cm]{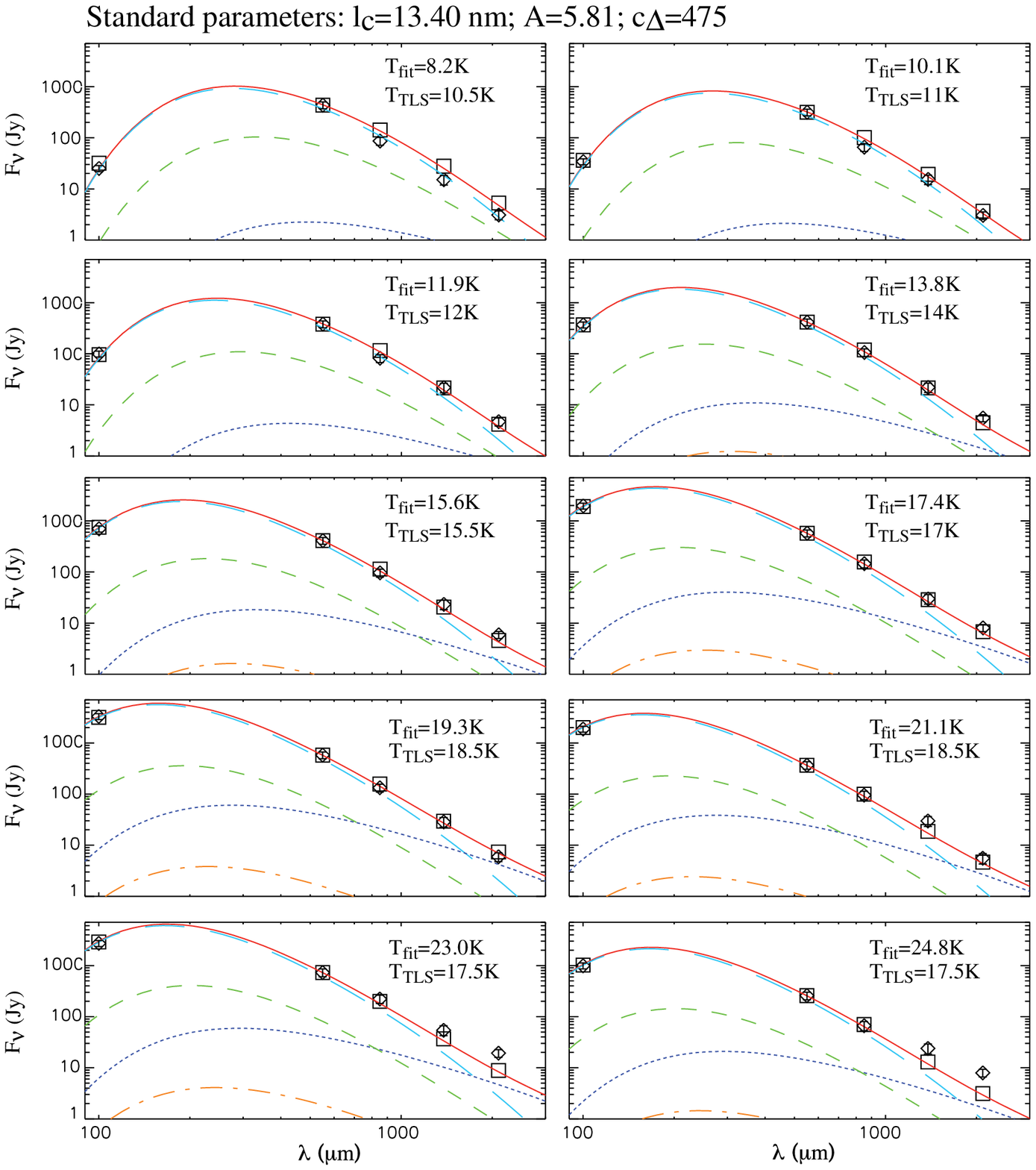}}
\end{center}
\caption{Continued.}
\end{figure*}

\subsubsection{Modeling of the Archeops spectra}
\label{sec_arch}
In this section, we perform the minimization on the Archeops spectra,
averaged in ten temperature bins. The IRAS 100 $\mic$ data were
also included in the spectra. Since the wavelength domain is limited
to $\lambda \simeq$2 mm, no free-free or spinning dust emission were
included in the fits. For each set of parameters [$l_c$, $A$,
$c_{\Delta}$], the model is computed for temperatures between 5 K and
35 K, with a step of 0.5 K, and adjusted to the spectra for the best
dust temperature for each spectrum, by looking for the lowest $\chi^2$
value. Therefore,
we determine the best set of parameters that is able to reproduce
all Archeops spectra by only changing the dust
temperature. Results of the minimization are presented in Figure
\ref{fig_arch} and Table \ref{tab_bootstrap}. A comparison between the
temperature derived from a modified black-body fit \citep{Desert08} and the
temperature deduced from the \TLS model are also given in Figure
\ref{fig_arch}. Whereas a modified black-body fit samples the range 8.2-24.8 K,
the \TLS model shows temperatures from 10 to 17.5 K. This
discrepancy is essentially due to the mm excess which can bias the
temperature determination when using a modified black-body fit. In Section \ref{sec_t},
we discuss the temperature estimates depending on the method used.

The parameters are better constrained than in Section \ref{sec_firas}, with
lower 1-$\sigma$ uncertainties, especially for $l_c$ with
a value of 5.11$\pm$0.09 nm. In Figure \ref{fig_arch} we can see that the
flattening of the spectra with increasing dust temperature (from top
to bottom panels, and from left to right) is mostly consistent
with the hopping relaxation process. However, the spectra with the
lowest dust temperature indicate that the Archeops data at longer
wavelengths and at 2 mm in particular, show a resonant absorption 
on the same order of magnitude or higher than implied by the hopping
relaxation process. Also, we note that the tunneling relaxation
contribution is much lower than the other processes. The minimization
of Galactic compact sources favors a lower intensity of the
TLS resonant process which results in the increase in the amplitude of the hopping
relaxation, as compared to the modeling of the FIRAS/WMAP
spectrum. In other words, the decrease in the $A$ parameter requires a higher value of $c_{\Delta}$.

\begin{table*}[!t] 
\caption[]{Parameters of the \TLS model. \label{tab_bootstrap}} 
\begin{center}
\begin{tabular}{l c c c c c c}
 \hline
\hline
  & \parbox{0.9cm}{\center T$\rm _d$ (K)} & \parbox{0.9cm}{\center $l_c$ (nm)}
  & \parbox{0.9cm}{\center $A$} & \parbox{0.8cm}{\center $c_{\Delta}$}
  & \parbox{0.8cm}{\center $\alpha$} & \parbox{1.5cm}{\center
    reduced $\chi^2$} \\
  \hline
Galactic diffuse emission (FIRAS/WMAP) & & & & & &\\
\hline
Parameters & 17.53& 23.05 & 9.38 & 242  & 1.45 & 1.95 \\  
1-$\sigma$ & $\pm $ 0.02 & $\pm $ 22.70 &$\pm $ 1.38 & $\pm$ 123 & $\pm $ 0.15 & -- \\
\hline
Galactic compact sources emission (Archeops) & & & & & &\\
\hline
Parameters & -- & 5.11 & 3.86 & 1333  & -- & 1.45\\   
1-$\sigma $ & -- & $\pm$ 0.09 & $\pm$ 0.13 & $\pm $ 68 & -- & -- \\
\hline
Galactic diffuse and compact sources emission  & & & & & & \\  
\hline
 Standard parameters & 17.26 & 13.40 & 5.81 & 475 & 1.55 & 2.53 \\
1-$\sigma $ & $\pm $ 0.02 &$ \pm $ 1.49 & $\pm $ 0.09 & $\pm $ 20 & $\pm $ 0.11 &  -- \\
  \hline
\end{tabular}
\end{center}
\end{table*}

\subsubsection{Modeling of the FIRAS/WMAP and Archeops spectra}
To derive general properties of Galactic dust, we performed a combined
fit of the diffuse medium and compact sources SEDs. Best-fit
parameters, denoted ``standard'' parameters are given in Table
\ref{tab_bootstrap}. Figures \ref{fig_tls} and \ref{fig_arch} also show
the results of the modeling when using the standard parameters. Since the dust temperature is only derived from
the FIRAS/WMAP SED, we would expect a similar value as in the combined
fit. However, because the FIR spectral shape of the BG emission
also depends on $l_c$, changes in $l_c$ produce changes in
$T_d$, since the two parameters are degenerate. We note a decrease in
$l_c$ with a value of 13.4 nm, compared to the value derived using
only the diffuse medium emission ($l_c=23.05$ nm). This explains that
the best temperature estimate ($T_d=17.26$ K) does not fully match the
value obtained when fitting only the FIRAS/WMAP data ($T_d=17.53$
K). Actually, the $l_c$, $A$, and $c_{\Delta}$ parameter values are
intermediate, compared to the values obtained when fitting
independently the diffuse medium emission on one side and compact
sources on the other. We note that using the same set of \TLS
  parameters for all Archeops compact sources does not reproduce the
  long wavelength data well in the case of the warmest
  spectra as determined by a modified black-body fit ($T_{fit}$=23.0-24.8 K.). In that case, the \TLS model underestimates the brightness at 1382 and 2098 $\mic$.

In Figure \ref{fig_tls} we do not see any significant difference
  in the modeling of the FIRAS/WMAP spectrum when using the best-fit
  or standard parameters. However, when looking at Figure
  \ref{fig_arch}, the use of the standard parameters has some non-negligible effects on the fit modeling. For instance, the fitting of
  the coldest spectrum shows an overestimate of the brightness at
  long wavelengths, which was not the case with the best-fit
  parameters. Also, the warmest spectra are described less well when adopting the standard
  parameters. This could indicate that dust properties are different
  in the diffuse medium observed with FIRAS/WMAP and in compact
  sources observed with Archeops.

\subsection{Discussion of fit results}
Parameters giving the best match between the \TLS model and the
Galactic emission spectra correspond to a dust temperature of $T_d=$17.26
K, a correlation length close to $l_c=$13 nm, an intensity of the TLS
processes with respect to the DCD one of about $A=$6, and an intensity parameter
for the hopping relaxation of $c_\Delta=$475. The dust temperature characterizing
the diffuse interstellar medium is quite close to the value of 17.5 K
derived by \citet{Boulanger96} at high Galactic latitudes.
We note that the
correlation length is within the same order of magnitude as the minimal
size of the smallest grains in the BG population. The largest grains
could result from aggregation of grains with a size close to the
correlation length.
The value of the
$A$ parameter is reasonable considering the paucity of
quantitative experimental data in this field
\citep[essentially][]{Bosch77}.
In solid-state physics, the phenomenology of these effects is
different, and these effects could result from distinct types of
disorder in the material. The moderate value found for the parameter
$A$ indicates that the relative intensity of the DCD and TLS processes
has the same order of magnitude as in the case studied by
\citet{Bosch77}. 

For the physical properties related to 
defects in the grain material, the measured values of defect
concentration and induced defects can easily vary over several orders
of magnitude. Therefore the derived $A$ value suggests a similar
origin for the DCD and TLS processes, and therefore justifies the use of
a similar density of state. The value of $c_{\Delta}$ is significantly
higher than the minimum value of 5.8 deduced from laboratory
experiments, at low temperature. Once again, in a material with a known
chemical composition, the magnitude of the physical effects caused by
defects in the material are likely to vary over several decades. The value derived from
observational data is in that sense fully acceptable. This value
indicates that the hopping relaxation is more efficient than the two
other TLS processes, related to tunneling and therefore highly
dependent on the distance between the two potential minima of
the TLS. However, we note that each TLS process is characterized by
its own emission profile, dependent on the parameters for each
site. For instance, modifying $\rm V_m$, $\rm V_0$, and $\rm V_{min} $
would induce variations in the emission profile of the hopping
relaxation effect. As a consequence the $c_{\Delta}$ value would 
have to be modified.

\section{Model predictions}
\subsection{Spectral shape evolution with temperature -
Emissivity spectral index} 
\label{sec_beta_predictions}
\begin{figure} 
\begin{center}
\includegraphics[width=8cm]{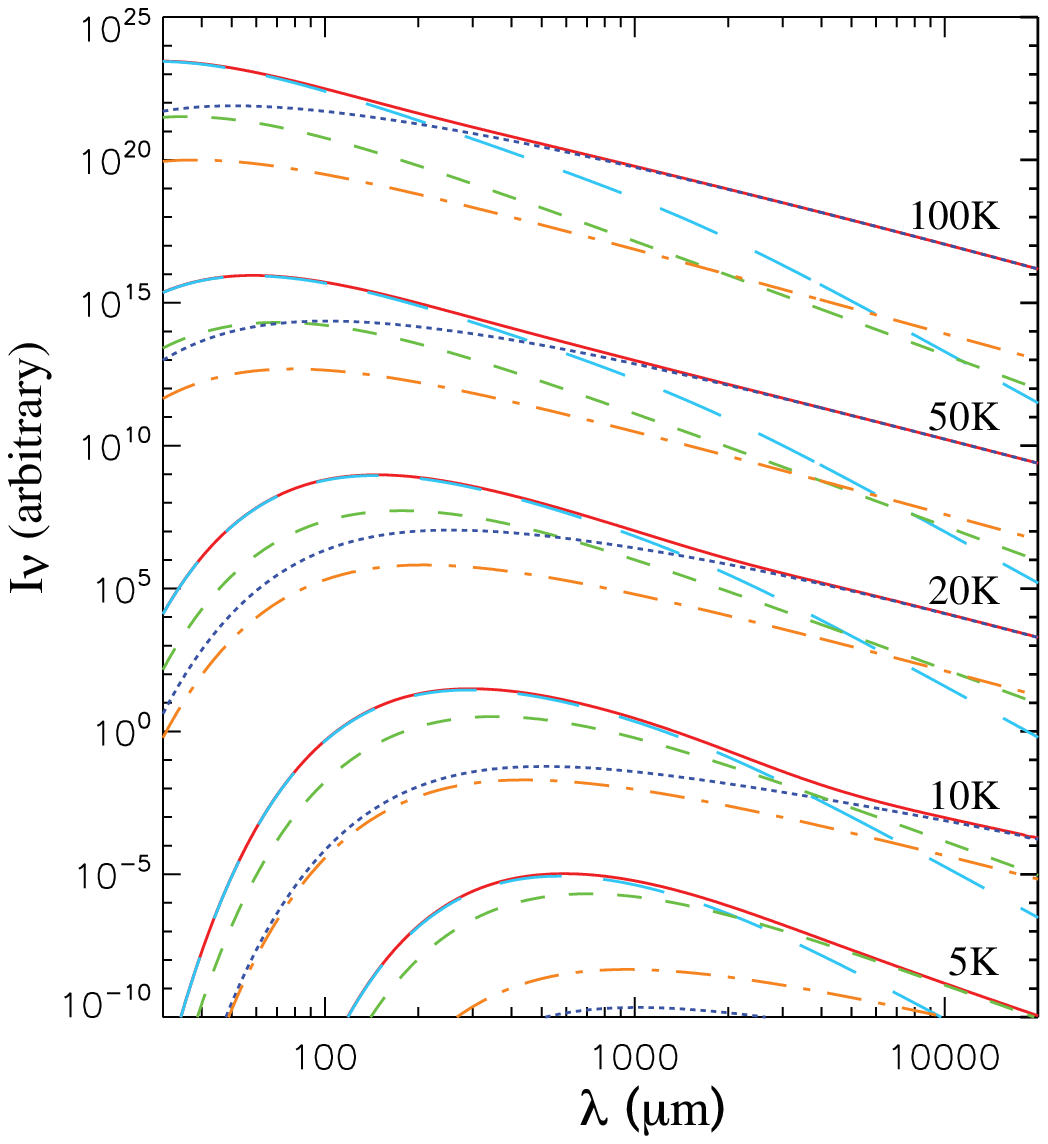}
\end{center} 
\caption{Prediction of the FIR/mm emission at different temperatures,
  deduced from the \TLS model using standard parameters (see Table \ref{tab_bootstrap}). The contribution of the different processes is given
  in the caption of Figure \ref{fig_tls}. \label{fig_evot}}
\end{figure}

In Figure \ref{fig_evot}, we present the predicted model SED evolution
using the standard parameters taken from Table \ref{tab_bootstrap} for different dust temperatures. The resonant absorption and the hopping
relaxation process have different behaviors with temperature: the
hopping relaxation increases with increasing temperature,
whereas the resonant absorption decreases. The tunneling relaxation
can be neglected over the range of temperature explored here,
since it appears to be systematically dominated by one of the other
TLS processes, and does not induce any specific variations in the dust
emission spectrum.  The millimeter excess emission, as well as the
flattening of the spectrum with temperature, is mainly produced by the
hopping relaxation. As a consequence, the excess increases with
temperature, and the transition between the DCD and TLS processes
shifts towards shorter wavelengths with increasing temperature
since the intensity of the hopping relaxation increases. 

\begin{figure} [!t]
\begin{center}
\includegraphics[width=8cm]{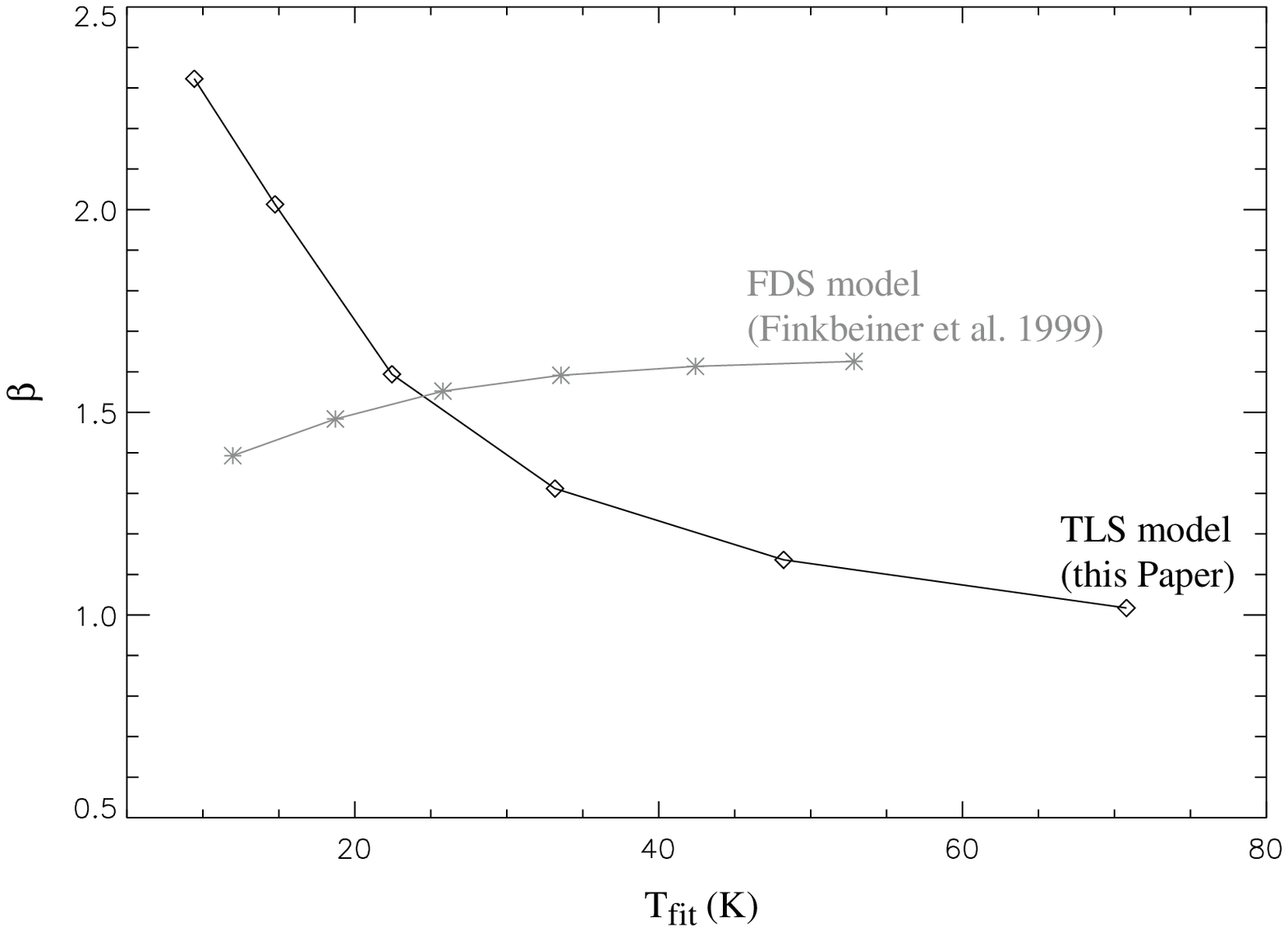}
\end{center} 
\caption{Prediction of the emissivity spectral index as a function of
  temperature, deduced from a modified black-body fit on emission
  spectra at various temperatures, in  the IRAS 100 $\mic$ and Archeops bands. The emission spectra have been
  generated using the \TLS (black) (using standard parameters, see
  Table \ref{tab_bootstrap}) and FDS (gray) models. \label{fig_t_beta_arch}}
\end{figure}

One of the most important aspects of the \TLS model to be explored is the variation
in $\beta$ with temperature. We compared the \TLS to the FDS model number 7 \citep[see Table 2
in][]{Finkbeiner99}. The same method has been applied to each model. We generated a set of emission
spectra for temperatures between 10 and 35 K corresponding to
  temperatures derived using the \TLS model (hereafter $\rm T_{\TLS}$)
  in one case, and to the warm component
(T$_2$) of the FDS model, which is related to that of the cold
component \citep[see Eq. 14 in][]{Finkbeiner99}, in the other. Each
spectrum has been integrated in
the IRAS 100 $\mic$ and Archeops photometric bands to produce simulated SEDs. We then fitted these fake SEDs at
different temperatures by a single
modified black body ($\rm T_d$-$\beta$ fit), assuming the same
absolute uncertainties as for the actual IRAS and Archeops data (see Section \ref{sec_cs}). Results of the $\rm T_d$-$\beta$ fits are presented
in Figure \ref{fig_t_beta_arch}. The \TLS model clearly
highlights an inverse relationship between the spectral index and the
temperature of the fit (hereafter $\rm T_{fit}$) in
agreement with the dust behavior in \citet{Desert08},
whereas the FDS model shows the opposite behavior, i.e. an increase
in $\beta$ with temperature. We also note that the FDS model is not able to predict $\beta$
values higher than 1.6 in the wavelength range observed with
Archeops. Using standard
parameters, the \TLS model predicts $\beta$ variation between 2.3 and
1 in the wavelength domain going from 100 $\mic$ to 2 mm, for $\rm
T_{fit}$ between 10 and 70 K. 

Models such as the FDS model, which
do not exhibit any intrinsic variation in the apparent spectral index
with temperature may still explain the obervations in principle,
provided an adequate temperature mixing along th LOS or within the
beam sufficiently alters the emitted SED. 

First, we note that, since
temperature mixing can only broaden the emitted SED, the intrinsic
$\beta$ should be larger than the range of observed ones. Since many
of the Archeops point sources and Planck cold cores \citep{Montier11} have
apparent $\beta$ larger than 3-4, this would require a model with a
very steep intrinsic $\beta$ value. Apart from the \TLS model proposed
here, which can produce this steep behavior under specific
circumstances, there has been no proposal for dust models that allows
such high $\beta$ values. 

Second, the effect of temperature mixing has been analyzed in several
studies. \citet{Malinen11}, for instance, have investigated this effect 
through 3D MHD simulation, taking both full
radiative transfer and a dust model with no intrinsic $\beta$
variations into account. They considered both cases, with and without embedded
stars heating the surrounding ISM, assuming reasonable luminosity and
spatial distribution for the sources. \citet{Paradis09} studied
the emissivity variations induced by an ISRF strength mixture using
the \citet{Dale01} concept of a power-law distribution of dust mass
subjected to a given heating intensity ($dM_d(X_{ISRF} \propto
X_{ISRF}^{-\alpha}dX_{ISRF}$). The $\alpha$ parameter has been tuned
in the range $\alpha$=1 (active star-forming regions) up to $\alpha$=2.5
(diffuse medium). Both methods give similar results. 

For cases with no
embedded source, temperature mixing can reproduce neither strong
wavelength-dependent variations in the emissivity nor the spread of
$\beta$ values observed. Indeed, \citet{Malinen11} show that the
model corresponding to that case actually produces a weak correlation
between $\rm T_d$ and $\beta$, which is the opposite of the
anticorrelation that is observed. 
In cases of star-forming regions, when discrete stellar heating is
included in the \citet{Malinen11} model, most LOS in their simulation
show a similar correlation trend. However, those LOS towards discrete
objects \citep[which corresponds to Dale's $\alpha$-parameter
$\simeq$1.25 in ][]{Paradis09} present a wavelength-dependent
variation in emissivity, leading to an apparent $\rm T_d$-$\beta$ anticorrelation
with a shape similar to what is observed toward some regions. Those map pixels
affected by temperature-mixing around heating sources, however, have
apparent temperatures higher than 14 K following \citet{Malinen11} and, as already
pointed out, $\beta$ values always lower than the ones
characterizing the SEDs without temperature mixing. 

As shown in Table \ref{tab_arch}, most of the Archeops point sources considered
here have temperatures lower than 14 K, down to 8 K, and high
spectral index values. This shows that they are most likely not affected by
temperature mixing and are actually closer to cold cores, as detected
in the Planck data \citep{Montier11}. It is there observed that such
sources exhibit narrow SEDs. For a single dust temperature along the
LOS, their narrowness is only compatible with spectral index values
higher than 3. And a strong distribution of temperature would require
still higher spectra index values to fit the observed SEDs.
For  the above reasons, we do not favor a model without intrinsic
$\beta$ variations and temperature mixing along the LOS to simultaneously explain
the diffuse MW SED and the Archeops point sources, and propose
that the \TLS model with the parameters derived here could reasonably
explain both. \\

\begin{figure} [!t]
\begin{center}
\includegraphics[width=8cm]{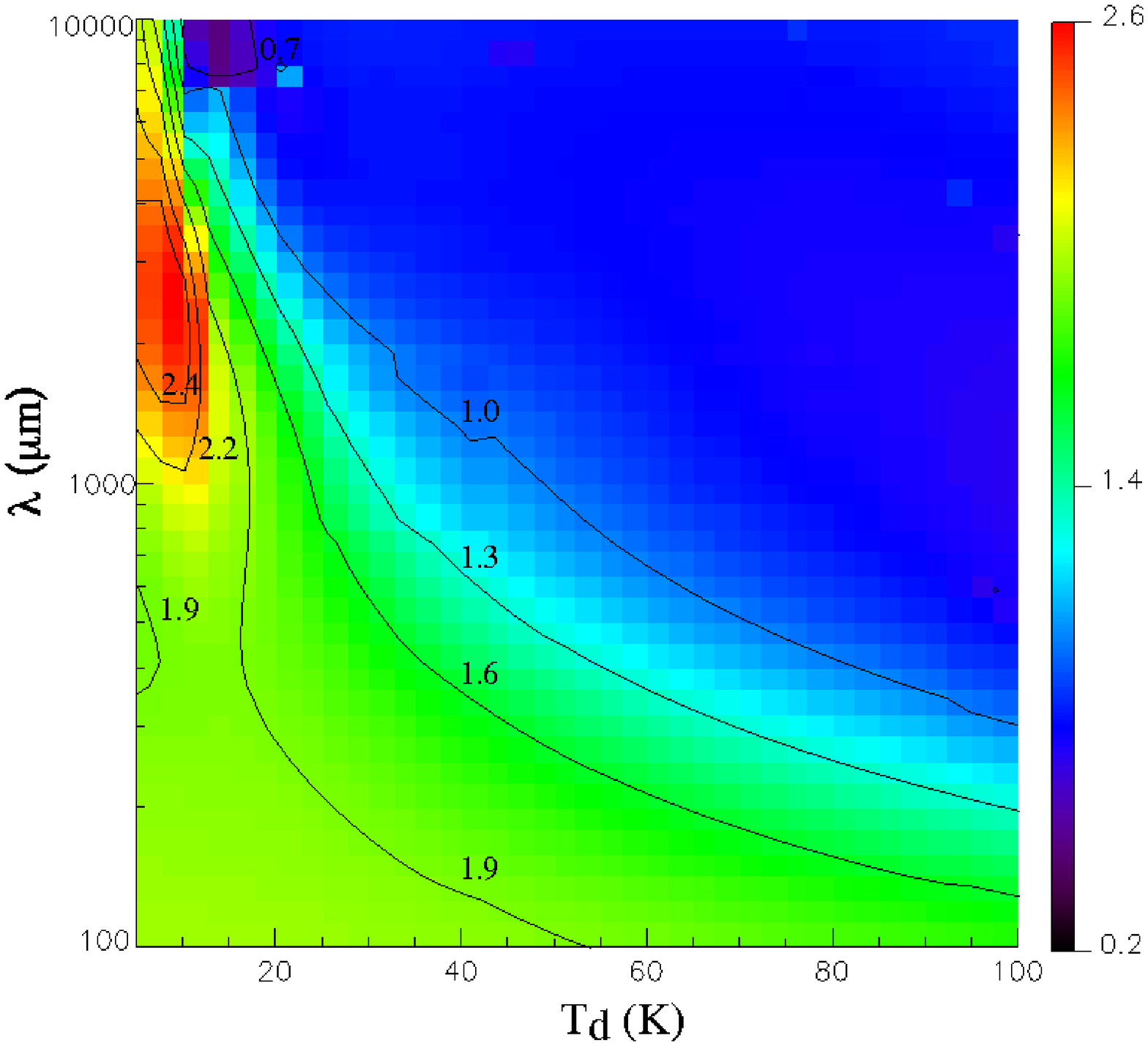}
\end{center} 
\caption{Prediction of the emissivity spectral index deduced from the
  \TLS model using standard parameters (see Table \ref{tab_bootstrap}), as a function of wavelength and temperature, with
  overplotted contours.  \label{fig_beta}}
\end{figure}

To extend the analysis on the $\beta$ variations, we present in Figure \ref{fig_beta} how the predicted $\beta$ changes as
a function of both temperature and wavelength, computed as $dln
  \epsilon_e/dln \lambda$. For temperatures below 15 K,
the spectral index is expected to exceed 2 and reach 2.6 at T$\rm
_d$=10 K, around 2 mm. Then $\beta$ decreases with increasing wavelength and gets as low as 0.5
at longer wavelengths, where dust emission is usually dominated by
other emissions such as free-free and synchrotron. Temperatures higher
than 20 K and 100 K lead to $\beta$ values less than 2 and 1.7 at $\lambda>100$ $\mic$, respectively. We note that at 2 mm,
in the range 10-40 K, the spectral index roughly varies from 2.6 to
1. To summarize, the $\beta$ values predicted by the \TLS model
vary strongly both with wavelength and dust temperature.

\begin{figure}
\begin{center}
\includegraphics[width=8cm]{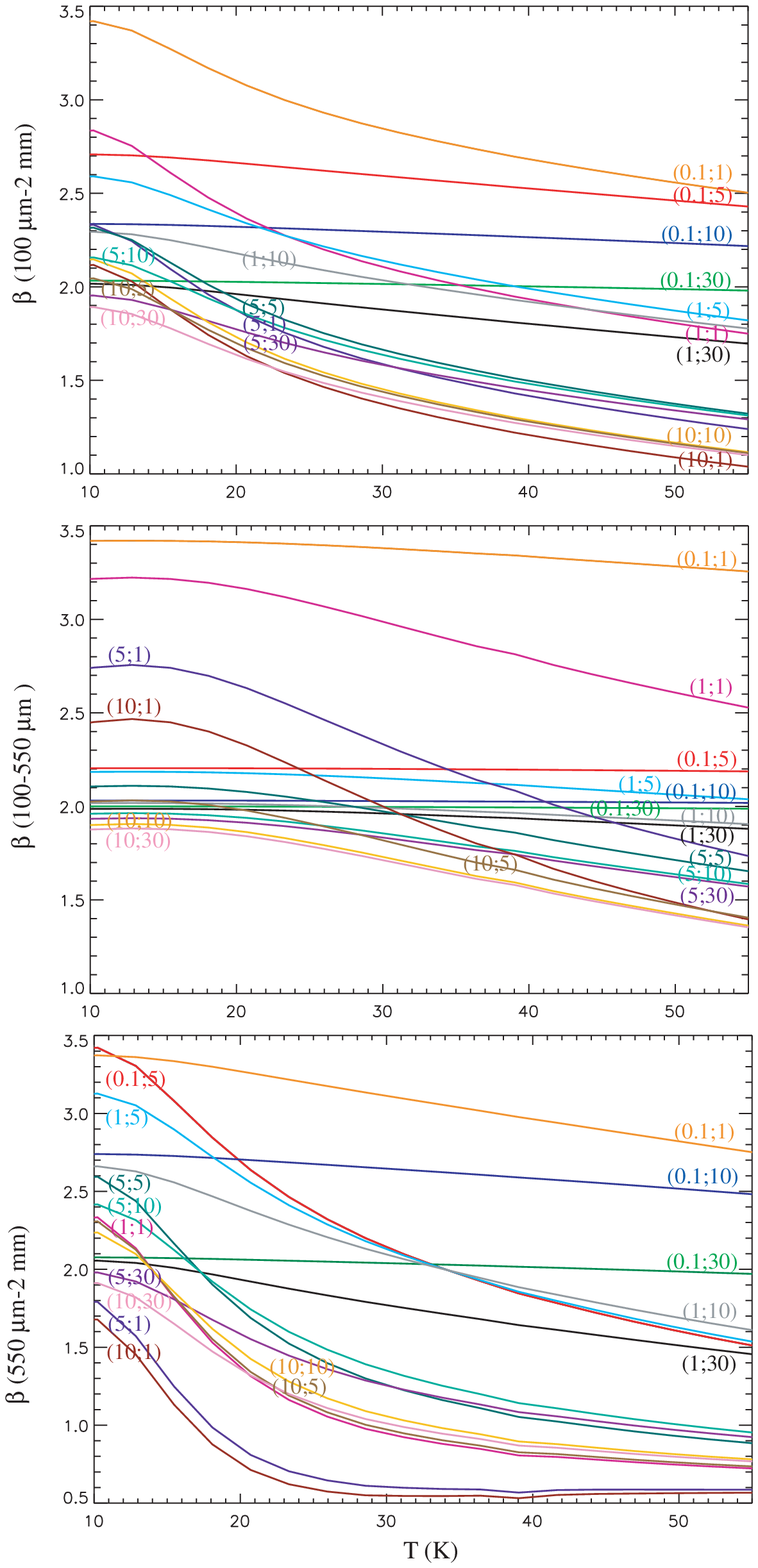}
\end{center}
\caption{Prediction of the emissivity spectral index in the range [100
  $\mic$-2 mm] (top panel), [100-550 $\mic$] (middle panel), [550
  $\mic$-2 mm] (bottom panel), deduced from the
  \TLS model as a function of dust temperature, for different
  parameter couples [$A$, $l_c$ (nm)] assuming $\rm c_{\Delta}=475$.  \label{fig_a_lc}}
\end{figure}

In Figure \ref{fig_a_lc} we show the evolution of $\beta$ in three
ranges of wavelengths ([100 $\mic$-2 mm], [100-550 $\mic$] and [550
$\mic$-2 mm]) for different couples of the parameters [$A$, $l_c$]. The
$c_{\Delta}$ value is fixed to the standard value of 475. For a given
$\beta$ and temperature, one can estimate the correlation length of
the dust grains, as well as the intensity of the TLS
effects. Regardless of the value of the parameters, Figure \ref{fig_a_lc} highlights a decrease in $\beta$ with temperature going from 10 to
  55 K for most [$A$, $l_c$] couples. The
$\beta$ variations are more pronounced as $A$ increases. 
The behavior of $\beta$ with temperature does not change
for $l_c\ge30$ nm, regardless of the $A$ value. In that case, the DCD part
has reached the asymptotic behavior in $\lambda^{-2}$, and only the TLS effects
can induce a decrease in $\beta$ with $T_d$. Low values of $l_c$
associated with low values of $A$ produce the
highest $\beta$ values, which can reach $\beta$=3.4-3.5 at low
temperatures in the three wavelength ranges. The DCD asymptotic
behavior in $\lambda^{-4}$ has not been reached in Figure \ref{fig_a_lc}, since
the TLS effects never vanish completely and $l_c$ values around 0.1 nm
would be required, which is not realistic since the interatomic distance is $\simeq$0.2-0.3 nm range. High values of $A$
induce low values of $\beta$ at high temperatures. The value of $\beta$ can be as
low as 0.55 between 550 $\mic$ and 2 mm, for temperatures higher than
25 K, in case of $A=10$ and $l_c=1$ nm. The same values of $A$ and
$l_c$, however, reproduce $\beta \simeq$ 1.4, for $T_d$=55 K in the
range 100-550 $\mic$. 

\subsection{Dust emissivity}
\begin{figure} 
\begin{center}
\includegraphics[width=8cm]{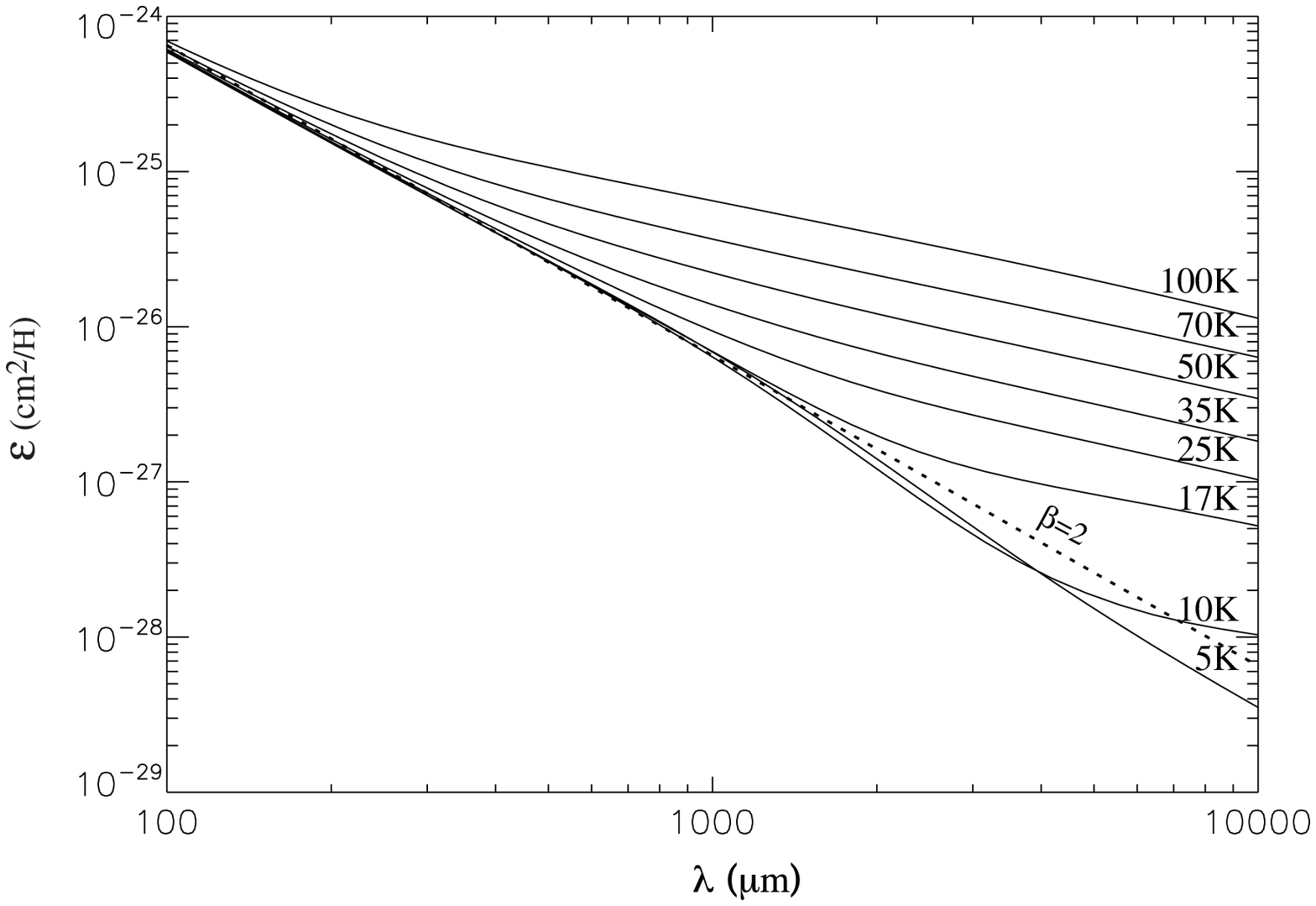} 
\end{center}
\caption{Predicted emissivity with the \TLS model using
the standard parameters given in Table \ref{tab_bootstrap}, as a function of
wavelength for different temperatures. The dashed line shows a $\lambda^{-2}$
emissivity power law for comparison. \label{fig_em}}
\end{figure}

One of the most important aspects of a given FIR/mm dust emission model
is to predict the dust emissivity and its variations with wavelength
and temperature. This point is pivotal both for inferring gas masses
and for separating foreground emission from the cosmic microwave
background (CMB).  Here, we quantify how adopting the predictions
of the \TLS model affects the dust mass determination obtained from
FIR and submm measurements. The reference emissivity
value is 1$\times10^{-25}$ cm$^2$/H at 250 $\mic$ in the solar
neighborhood \citep{Boulanger96}. This value has been derived for a
dust temperature of 17.5 K. Besides variations with wavelength, the \TLS model predicts variations with temperature. We
impose the model to recover the
Galactic emissivity reference at 250 $\mic$ from \citet{Boulanger96},
for a temperature of 17.5 K. Emissivity predictions have been integrated for
different temperatures in the IRAS, Herschel PACS and SPIRE, Planck HFI and
LFI channels. The color correction has then been applied to each instrument. Values of the predicted
emissivities are given in Table \ref{tab_em}.
Figure \ref{fig_em} shows the emissivity predictions for temperatures between
5 K and 100 K, overlaid with a $\lambda^{-2}$ power-law emissivity. We
note that between 5 and 17 K, the emissivity variations follow a
$\lambda^{-2}$ power law in the range 100 $\mic$-1 mm. At higher
temperatures, the emissivity spectra become flatter than a modified black-body
with $\beta$=2, starting in the FIR domain and flattening even
more with wavelength.  We observe that emissivities at 5 and 10
K only differ for $\lambda>4$ mm. For T$\rm _d>$10 K, we can see that
the difference between the curves increases significantly with
wavelength and temperature. Whereas the discrepancy between
predictions at 5 K and 100 K is about a factor 2 at 250 $\mic$, it is
about a factor 35 at 2 mm, and $\simeq$100 at 4 mm.

Therefore, when assuming no dependency of the emissivity spectrum with
either wavelength or temperature, one can significantly bias the mass
estimate, especially when derived from submm/mm data and extrapolated
to shorter wavelengths. Indeed, when assuming an emissivity decrease
in $\lambda^{-2}$ in the submm/mm domain, one would overestimate the
dust temperature and underestimate the emissivity in the FIR, and
therefore would overestimate the dust mass. For instance, this is the
case for those SMC molecular clouds whose masses have been derived from mm
data and found to be systematically higher (twice on average) than virial
masses \citep{Bot07}.

\begin{sidewaystable*}
\vspace{2cm}
\begin{center}
\resizebox{\textwidth}{!}{
\begin{tabular}{lcccccccccccccc}
\hline
\hline
$\lambda$ ($\mu$m) & 100 & 160 & 250 & 350 & 350 & 500 & 550 & 850 &
1382 & 2096 & 2998 & 4286 & 6818 & 10000 \\
Photometric bands & IRAS & PACS & SPIRE & SPIRE & HFI & SPIRE & HFI &
HFI & HFI & HFI & HFI & LFI &LFI &LFI \\
\hline
5 K&  6.51$\times10^{-25}$ &  2.55$\times10^{-25}$ &
1.06$\times10^{-25}$&   5.54$\times10^{-26}$  &5.36$\times10^{-26}$ &
2.85$\times10^{-26}$&   2.38$\times10^{-26}$ &   1.02$\times10^{-26}$&
3.64$\times10^{-27}$ & 1.23$\times10^{-27}$ &   5.44$\times10^{-28}$ &
2.24$\times10^{-28}$ &   7.84$\times10^{-29}$ &  2.88$\times10^{-29}$ \\
10 K& 6.51$\times10^{-25}$ &  2.55$\times10^{-25}$
&1.06$\times10^{-25}$& 5.53$\times10^{-26}$  & 5.35$\times10^{-26}$ & 2.81$\times10^{-26}$&   2.35$\times10^{-26}$ &   9.73$\times10^{-27}$&
3.27$\times10^{-27}$ & 1.06$\times10^{-27}$ &   4.83$\times10^{-28}$ &
2.35$\times10^{-28}$ &   1.33$\times10^{-28}$ &  9.25$\times10^{-29}$ \\
17 K& 6.52$\times10^{-25}$ &  2.56$\times10^{-25}$
&1.07$\times10^{-25}$& 5.56$\times10^{-26}$  & 5.38$\times10^{-26}$ & 2.83$\times10^{-26}$&   2.37$\times10^{-26}$ &   1.01$\times10^{-26}$&
3.95$\times10^{-27}$ & 1.82$\times10^{-27}$ &   1.24$\times10^{-27}$ &
9.24$\times10^{-28}$ &   6.77$\times10^{-28}$ &  4.56$\times10^{-28}$
\\
25 K& 6.55$\times10^{-25}$ &  2.58$\times10^{-25}$
&1.09$\times10^{-25}$& 5.78$\times10^{-26}$  & 5.59$\times10^{-26}$ & 3.07$\times10^{-26}$&   2.61$\times10^{-26}$ &   1.27$\times10^{-26}$&
6.26$\times10^{-27}$ & 3.64$\times10^{-27}$ &   2.72$\times10^{-27}$ &
2.03$\times10^{-27}$ &   1.40$\times10^{-27}$ &  9.01$\times10^{-28}$
\\
35 K& 6.62$\times10^{-25}$ &  2.64$\times10^{-25}$
&1.15$\times10^{-25}$& 6.35$\times10^{-26}$  & 6.15$\times10^{-26}$ & 3.63$\times10^{-26}$&   3.16$\times10^{-26}$ &   1.76$\times10^{-26}$&
1.01$\times10^{-26}$ & 6.33$\times10^{-27}$ &   4.84$\times10^{-27}$ &
3.61$\times10^{-27}$ &   2.48$\times10^{-27}$ &  1.60$\times10^{-27}$ \\
50 K& 6.76$\times10^{-25}$ &  2.78$\times10^{-25}$
&1.28$\times10^{-25}$& 7.64$\times10^{-26}$  & 7.43$\times10^{-26}$ & 4.81$\times10^{-26}$&   4.31$\times10^{-26}$ &   2.70$\times10^{-26}$&
1.71$\times10^{-26}$ & 1.13$\times10^{-26}$ &   8.85$\times10^{-27}$ &
6.68$\times10^{-27}$ &   4.65$\times10^{-27}$ &  3.01$\times10^{-27}$ \\
70 K& 7.03$\times10^{-25}$ &  3.04$\times10^{-25}$
&1.53$\times10^{-25}$& 9.98$\times10^{-26}$  & 9.73$\times10^{-26}$ & 6.88$\times10^{-26}$&   6.32$\times10^{-26}$ &   4.32$\times10^{-26}$&
2.92$\times10^{-26}$ & 2.01$\times10^{-26}$ &   1.60$\times10^{-26}$ &
1.22$\times10^{-26}$ &   8.54$\times10^{-27}$ &  5.55$\times10^{-27}$ \\
100 K& 7.57$\times10^{-25}$ &  3.57$\times10^{-25}$
&2.04$\times10^{-25}$& 1.46$\times10^{-25}$  & 1.43$\times10^{-25}$ & 1.09$\times10^{-25}$&   1.02$\times10^{-25}$ &  7.46$\times10^{-26}$&
5.28$\times10^{-26}$ & 3.72$\times10^{-26}$ &   2.97$\times10^{-26}$ &
2.26$\times10^{-26}$ &   1.56$\times10^{-26}$ &  9.87$\times10^{-27}$ \\
\hline
\end{tabular}}
\end{center}
\caption{Predicted emissivities in the IRAS, Herschel, and Planck photometric bands, deduced from the \TLS model with the standard parameters given in Table \ref{tab_bootstrap}, as a function of temperature. The emissivities have been normalized to $1\times 10^{-25}$ cm$^2$/H, at 250 $\mic$ for a
temperature of 17.5 K.\label{tab_em} Note: Table 5 is also available in machine-readable form in the electronic
edition of $A\&A$.}
\end{sidewaystable*}

\subsection{Temperature estimate}
\label{sec_t}
Dust temperature is often derived using a modified black-body fit with T$\rm _d$
and $\beta$ taken as free parameters, or with $\beta$ set to a fixed value. The latter temperature does not necessarily equal the physical
temperature obtained using the \TLS model, as already discussed in Section
\ref{sec_beta_predictions}. In particular, the
temperature deduced from a T$\rm _d$-$\beta$ fit is highly dependent
on the wavelength range considered. To compare the fitted temperature
with the physical one, we first generated several
emission spectra at different temperatures between 10 and 60 K using
the \TLS model. We then integrated the
\TLS model predictions in the IRAS 100 $\mic$ band (crucial for estimating the
dust temperature), Herschel PACS 160 $\mic$, and SPIRE 250, 350, and
500 $\mic$ bands. We then fitted each spectrum using a least-square fit
method. We also performed the analysis by integrating the \TLS
model in the IRAS 100 $\mic$ channel, and Planck HFI 350, 550, and 850
$\mic$ bands (857, 545, and 353 GHz), before fitting the spectrum with a modified black-body
model. For both cases (IRAS-Herschel bands and IRAS-Planck bands) we
have also made the test using a T$\rm _d$-$\beta$ fit with
$\beta=2$. For each band, we took the calibration uncertainties into account,
i.e. 13.5$\%$ for IRAS \citep[absolute uncertainty of the new
generation of IRAS maps][]{Miville05}, 20$\%$ for PACS, 15$\%$ for
SPIRE, and 7$\%$ for HFI.  The relationship between the dust
temperature derived using the \TLS model and
the T$\rm _d$-$\beta$ fit is modeled using a polynomial function, such as
\begin{equation}
\label{eq_t}
T_{\TLS}=\sum \gamma_nT_{fit}^n.
\end{equation}
The best polynomial coefficients derived from a fit are given in Table \ref{tab_t}.
\begin{table*} 
\begin{center}
\begin{tabular}{l c c c c c c c c}
 \hline
\hline
 Fit & $\lambda$ bands  & n & $\gamma_0$ & $\gamma_1$ & $\gamma_2$ &
$\gamma_3$ & $\gamma_4$ & $\gamma_5$  \\
  \hline
free T$\rm _d$, $\beta$ & [100-500] $\mic$ & 4 & -2.80348 & 1.38870 & -0.0141556
& 5.37193$\times 10^{-5}$ & -1.51050$\times 10^{-7}$ & --\\
free T$\rm _d$, $\beta$ & [100-850] $\mic$ & 5 & -0.328268 & 1.17842 &
-0.0117752 & 5.28944e-05 &  -8.77938e-08 & 2.81994e-11  \\
free T$\rm _d$, $\beta$=2 & [100-500] $\mic$ & 4 & 13.7673 & -1.99513 & 0.226583
& -0.00715176 & 8.32780e-05 & -- \\
free T$\rm _d$, $\beta$=2 & [100-850] $\mic$ & 5 & -17.8000 &  8.38173 &  -1.05454 &
0.0673555 & -0.00197337 & 2.18602e-05 \\
\hline
\end{tabular}
\end{center}
\caption{Polynomial coefficients (see Eq. \ref{eq_t}) relating the dust
temperature derived from the \TLS model and those derived from a
modified black-body fit. \label{tab_t}}
\end{table*}

\begin{figure}
\begin{center}
\includegraphics[width=8cm]{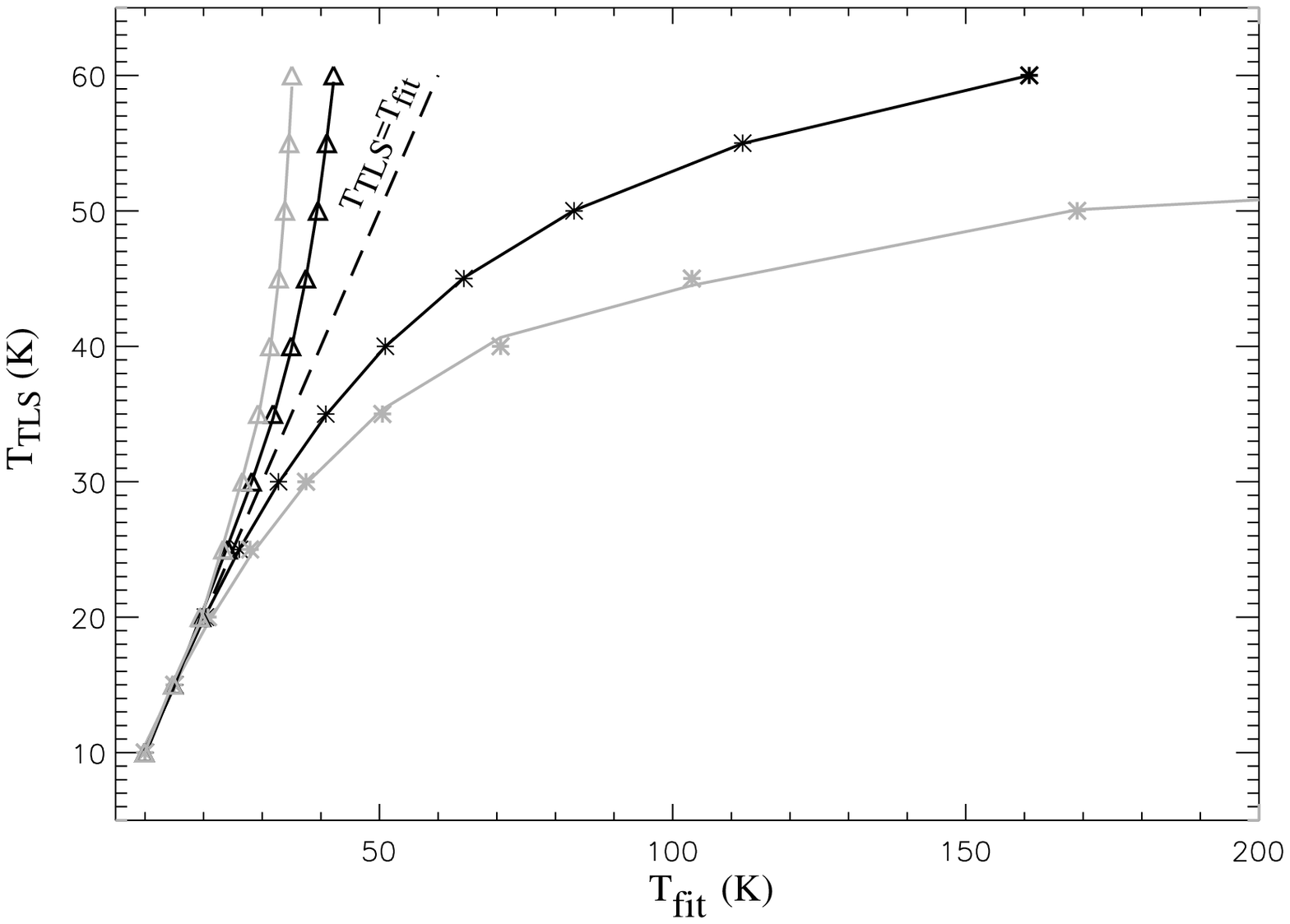}
\caption{
Temperature derived using the \TLS model ($\rm T_{\TLS}$) using the
standard parameters given in Table \ref{tab_bootstrap}, as a function
of temperature derived from a modified black-body fit with a free $\beta$
parameter ($\rm T_{fit}$) in the IRAS 100 $\mic$ - Herschel 160, 250,
350, and 500 $\mic$ channels (black stars), and in the IRAS 100 $\mic$
- Planck 350, 550, and 850 $\mic$ channels (gray stars).  The triangles
show the temperatures derived from a modified black-body fit with $\beta=2$,
using the IRAS-Herschel channels (black) and the IRAS-Planck channels
(gray). The dashed line corresponds to $\rm T_{fit}=T_{\TLS}$. The
continuous lines shows the best polynomial fits.\label{fig_temps}}
\end{center}
\end{figure}

The comparison between the temperature estimates is shown in
Figure \ref{fig_temps}. The behavior varies significantly when
deriving the temperature in the 100-500 $\mic$ range, or in the
100-850 $\mic$ range, and whether a free $\beta$ is
used. Up to 25 K, we observe good agreement between $\rm
T_{\TLS}$ and $\rm T_{fit}$, no matter which model is used, whereas it is
not the case for higher temperatures. When using a fixed (free)
$\beta$, $\rm T_{fit}$ is systematically lower (larger) than $\rm
T_{\TLS}$. Also, the discrepancy between the temperatures is greater
when enlarging the wavelength domain to derive $\rm T_{fit}$, and when
$\beta$ is left as a free parameter. This result highlights that the longest wavelengths ($\lambda>550$ $\mic$) should not
be used to determine the dust temperature from a single modified black-body
fit, since they are affected by significant submm/mm
emission excess in the \TLS model. In addition, one should be careful
when using a T$\rm _d$-$\beta$ fit, especially when $\beta$ is a free
parameter of the fit, and for $\rm T_{fit}>25$ K. In that case, 
Equation \ref{eq_t} and Table \ref{tab_t} should be used to derive 
the actual physical dust temperature from a temperature deduced from
a T$\rm _d$-$\beta$ fit. 

We note that if the temperature estimate is
performed on a wavelength range that extends up to 2 mm, as is the
case when using the Archeops data for instance, the discrepancy
between $\rm T_{fit}$ and $\rm T_{\TLS}$ is shifted, and roughly starts and
increases for temperatures over 18 K, instead of 25 K (see
temperatures presented in Figure \ref{fig_arch}). 

\section{Discussion}
One important aspect of dust emission is to be able to
subtract the foreground emission. According to the \TLS model,
emission from large ISM grains made of amorphous material is
significantly more complicated than predicted by a single
modified black-body emission with a single emissivity index
$\beta$. Actually, the model predicts emissivity index
variations as a function of both temperature and
wavelength. Therefore, within the framework of this model, it is
important to take into account both dependences for component
separation. Extrapolation of emission from the FIR to mm wavelengths
with a single modified black body with a constant $\beta$ could produce
a wrong estimate of dust emission in the CMB wavelength domain.  
The above variations are strictly related to the amorphous
nature of the material composing the grain, and should not be mistaken
with variations that can result, for instance, from modifications of
the grain structure, such as changes in the grain size or dust
aggregation \citep{Stepnik03, Paradis09}, which can occur in dense
regions of the ISM.

However, we caution the reader that the interpretation of the
model prediction may become inaccurate when high temperatures
($\simeq$100 K) are concerned. Indeed, in the current
version of the \TLS model, only the fundamental energy of the TLS
states is considered, and the model requires improvement by adding
 the effect of excited states in the two-level systems
\citep{Gromov11}.

The \TLS model predicts a wavelength and temperature dependence of
the spectral index that characterizes the emission and absorption
properties of the amorphous dust in the FIR and mm range. Coming from
solid states physics, it should apply with a high degree of
universality to all amorphous materials. Obviously, the exact profile
of the emission spectrum depends on parameters that define the
disorder, i.e. the amorphous structure of a material. Indeed, the
amorphous state is not well defined as crystals, and atoms can be
organized in a variety of local arrangements giving rise to a
diversity of structures on short and medium length scales. But the
emission profiles present some common characteristics: in the low-temperature range a spectral index value of the order of 2 in the FIR
range, followed by variations in the $\beta$ towards higher or lower
values at longer wavelengths. As the temperature increases, the
absorption is enhanced and is characterized by $\beta$ values that
decrease. These specificities can be observed in Figure
\ref{fig_beta}, for the given sets of parameters that fit the
observational data. However, this general behavior is also observed
in laboratory experiments performed on amorphous materials and grains
of astrophysical interest \citep{Agladze96,Mennella98,Boudet05}. In
particular, recent studies on three analogs of amorphous silicate
Mg$_2$SiO$_4$, MgSiO$_3$, and CaMgSi$_2$O$_6$ reveal that this
temperature and wavelength dependence of the absorption is observed on
all the samples, but disappear when the same samples are annealed
until crystallization \citep{Coupeaud11}.These analyses
indicate that the $\rm T_d$-$\beta$ variations are likely to result from intrinsic dust
properties that can be reproduced for the first time by a dust
emission model such as the \TLS one.

\section{Conclusions}
The theoretical \TLS model described in a previous work (Paper I)
and based on the amorphous structure of grains was compared to
astrophysical data, such as the FIRAS/WMAP and Archeops data. The
FIRAS/WMAP spectrum represents the diffuse interstellar medium in the
stellar neighborhood, whereas the Archeops data characterize dust
properties in a variety of compact sources, where a significant
inverse relationship between the dust temperature and the emissivity
spectral index had been proven. We performed a $\chi^2$
minimization to determine standard values for the parameters of the
\TLS model selected to capture the spectral and temperature variations
in the model. These free parameters are the dust temperature ($T_d$), the
charge correlation length ($l_c$) that controls the wavelength where the
inflection point between the two $\beta=2$ and $\beta=4$ ranges
occurs, the intensity factor of the TLS processes ($A$) with respect to the
DCD effect, and the intensity factor of the TLS/hopping
process ($c_{\Delta}$). Results indicate that emission in the submm/mm is dominated by
the hopping relaxation. According to the model, the BG emission in the
FIR/mm domain depends on wavelength and temperature, which is
fundamental both for dust mass determination from FIR/submm
measurements but also for component separation.

Using the best-fit parameters ($T_d=17.26$ K, $l_c$=13.4 nm, $A$=5.81, and
$c_{\Delta}$=475) allowing reproduction of both the emission
from the diffuse medium and the compact sources, the model predicts significant
$\beta$ variations for temperatures between 5 and 100 K, with a maximum value of 2.6 at
2 mm. The \TLS model is presently the only model able to predict $\beta$
variations with temperature and wavelengths, as observed in both
observational and laboratory data. The dust emissivity can be seriously
underestimated if its variations with temperature and wavelength are
not taken properly into account, generally inducing overestimates of the dust mass.
Similarly, the comparison between temperatures derived from the \TLS
model ($\rm T_{\TLS}$)
and those derived from a T$\rm _d$-$\beta$ fit ($\rm T_{fit}$),
between 100 and 500 $\mic$, or 100 and 850 $\mic$, are in good agreement for
temperatures below 25 K. Above this value, the derived temperatures
differ significantly from each other, in particular when emission far from the emission peak is
considered in the fit. Indeed, a modified black-body fit is too simplistic to
reproduce the mm emission excess, which increases with
temperature. However, our study enables the possibility of deducing $\rm
T_{\TLS}$ from a given $\rm T_{fit}$. We also predicted dust
emissivities in the IRAS 100 $\mic$, Herschel, and Planck bands for
$\rm T_{\TLS}$ between 5 and 100 K, which are useful for comparison
with the Planck and Herschel data. For instance, the provided tables
and figures  can be
used to derive the charge correlation length and the intensity of the TLS processes
from the Herschel and Planck data.

The \TLS model is the first physically-motivated model able to
reproduce both diffuse medium and compact sources, observed with
FIRAS/WMAP and Archeops. In the
future, the Planck data will be used to systematically compare the model prediction
with submm data.

\begin{acknowledgements}
We are very grateful to Bruce Draine for being the referee and for all his constructive comments, which greatly
improved the manuscript.
D. P. acknowledges grant support from the Centre National d'Etudes
Spatiales (CNES). J.-Ph. B. acknowledges the Lorentz Center
Workshop on "Herschel and the Characteristics of Dust in Galaxies",
which prompted useful discussions of the subject. V. G. acknowledges support from CNRS red positions during part of his work
on this paper.
\end{acknowledgements}

\end{document}